\documentclass[fleqn,usenatbib]{mnras}
\usepackage{mathptmx}
\usepackage[T1]{fontenc}
\usepackage{ae,aecompl}

\usepackage{graphicx}
\usepackage{amsmath}
\usepackage{amssymb}
\usepackage{bm}
\usepackage{pdflscape}	
\usepackage{color}
\usepackage{xcolor}
\usepackage{exscale}
\usepackage{natbib}
\usepackage{rotating}
\usepackage{rotate}
\usepackage{afterpage}
\usepackage{booktabs,threeparttable}
\usepackage{relsize}
\usepackage{lscape}
\usepackage{tabularx}
\usepackage{longtable}
\usepackage{ltxtable}
\usepackage{txfonts}
\usepackage{bbm}
\usepackage{multirow}
\usepackage{xurl}
\usepackage{hyperref}
\usepackage{soul}
\usepackage{units}

\newcommand{\ie}{{i.e.}}

\newcommand{\eg}{{e.g.}}

\newcommand{\py}{\small\ttfamily}

\newcommand{\ee}{\mathrm{e}}
\newcommand{\bv}[1]{\mathbfit{#1}}
\newcommand{\bmat}[1]{\mathbfss{#1}}
\newcommand{\x}{$\times$}
\newcommand{\btop}{{\bm{\top}}}
\newcommand{\bdot}{{\bm{\cdot}}}

\newcommand{\planck}{\textit{Planck} }

\title[10x2pt Modelling]{Cosmology from weak lensing, galaxy clustering, CMB lensing and tSZ:\\I. 10\x 2pt Modelling Methodology}

\author[X. Fang et al.]{Xiao Fang\thanks{E-mail: xfang@berkeley.edu},$^{1,2}$ Elisabeth Krause,$^{2,3}$ Tim Eifler,$^{2}$ Simone Ferraro$,^{4,1}$ Karim Benabed,$^{5}$ Pranjal R. S.,$^{2}$
\newauthor
Emma Ayçoberry,$^{5}$ Yohan Dubois,$^{5}$ \& Vivian Miranda$^{6}$
\\
$^{1}$Berkeley Center for Cosmological Physics, UC Berkeley, CA 94720, USA\\
$^{2}$Department of Astronomy and Steward Observatory, University of Arizona, 933 North Cherry Avenue, Tucson, AZ 85721, USA\\
$^{3}$Department of Physics, University of Arizona, 1118 E. Fourth Street, Tucson, AZ 85721, USA\\
$^{4}$Lawrence Berkeley National Laboratory, One Cyclotron Road, Berkeley, CA 94720, USA\\
$^{5}$Sorbonne Université, CNRS, UMR7095, Institut d’Astrophysique de Paris, 98 bis Boulevard Arago, F-75014, Paris, France\\
$^{6}$C. N. Yang Institute for Theoretical Physics, Stony Brook University, Stony Brook, NY 11794
}

\begin{document}

\date{Accepted . Received ; in original form }

\pagerange{\pageref{firstpage}--\pageref{lastpage}} \pubyear{2023}

\maketitle

\label{firstpage}

\begin{abstract} 
The overlap of galaxy surveys and CMB experiments presents an ideal opportunity for joint cosmological dataset analyses. In this paper we develop a halo-model-based method for the first joint analysis combining these two experiments using 10 correlated two-point functions (10\x2pt) derived from galaxy position, galaxy shear, CMB lensing convergence, and Compton-$y$ fields. We explore this method using the Vera Rubin Observatory Legacy Survey of Space and Time (LSST) and the Simons Observatory (SO) as examples. We find such LSS\x CMB joint analyses lead to significant improvement in Figure-of-Merit of $\Omega_m$ and $S_8$ over the constraints from using LSS-only probes within $\Lambda$CDM. We identify that the shear-$y$ and $y$-$y$ correlations are the most valuable additions when tSZ is included. We further identify the dominant sources of halo model uncertainties in the small-scale modelling, and investigate the impact of halo self-calibration due to the inclusion of small-scale tSZ information.
\end{abstract}

\begin{keywords}
cosmological parameters -- cosmic background radiation -- large-scale structure of Universe -- cosmology: theory
\end{keywords}

\renewcommand{\thefootnote}{\arabic{footnote}}
\setcounter{footnote}{0}

\section{Introduction}
\label{sec:intro}
Observations of the Universe's large-scale structure (LSS) provide valuable insights into cosmic structure formation and expansion history, enabling tests of theories of gravity and constraints on the mass and number of neutrino species, the nature of dark matter, and dark energy. To address these science questions, several galaxy survey experiments have been developed, including the Kilo-Degree Survey \citep[KiDS,][]{2017MNRAS.465.1454H,2021A&A...646A.140H}, the Dark Energy Survey \citep[DES,][]{2018PhRvD..98d3526A,2022PhRvD.105b3520A}, the Hyper Suprime-Cam \citep[HSC,][]{2019PASJ...71...43H}, the Dark Energy Spectroscopic Instrument \citep[DESI,][]{2016arXiv161100036D}, the Vera C. Rubin Observatory Legacy Survey of Space and Time \citep[LSST,][]{2019ApJ...873..111I}, the Nancy Grace Roman Space Telescope \citep{2019arXiv190205569A}, the Euclid mission \citep{Euclid_WhitePaper}, and the Spectro-Photometer for the History of the Universe,
Epoch of Reionization, and Ices Explorer \citep[SPHEREx,][]{2014arXiv1412.4872D}. Jointly analysing multiple cosmological probes, such as ``3\x2pt'' analyses that combine galaxy clustering and weak lensing statistics, has become the standard in KiDS, DES, and HSC since it increases the overall constraining power and the robustness to systematic effects. This ``multi-probe analysis'' approach is expected to remain a key strategy in extracting cosmological information from upcoming galaxy survey experiments.

A series of experiments have been dedicated to measuring the energy composition and structure growth of the Universe at much higher redshifts via the cosmic microwave background (CMB). Whether there exists an $S_8$ parameter difference within the standard $\Lambda$CDM model between high-$z$ CMB measurements from \planck \citep{2020A&A...641A...6P} and several low-$z$ measurements, such as KiDS \citep{2018MNRAS.474.4894J,2021A&A...646A.140H}, DES \citep{2018PhRvD..98d3526A,2019PhRvD.100b3541A,2022PhRvD.105b3520A}, unWISE and \planck CMB lensing tomography \citep{2020JCAP...05..047K, 2021JCAP...12..028K}, and Atacama Cosmology Telescope (ACT) CMB lensing \citep{2023arXiv230405202Q,2023arXiv230405203M}, has been extensively investigated, yet the issue remains unresolved. The significance of the tension/agreement between these measurements, and potential deviations from the $\Lambda$CDM model, will become more clear with future CMB analyses from ACT, \citep{2020JCAP...12..047A,2020JCAP...12..045C}, the South Pole Telescope \citep[SPT,][]{2014SPIE.9153E..1PB,2021PhRvD.104b2003D,2022arXiv221205642B}, the Simons Observatory \citep[SO,][]{2018SPIE10708E..04G,2019JCAP...02..056A}, and the CMB Stage-4 survey \citep[S4,][]{2016arXiv161002743A,2019arXiv190704473A}, along with wider and deeper galaxy surveys.

CMB secondary effects, such as CMB lensing \citep{2006PhR...429....1L} and thermal Sunyaev-Zel'dolvich (tSZ) effect \citep{2002ARA&A..40..643C}, trace the foreground gravitational potential and baryon distributions, hence are complementary probes of LSS. As the sensitivities of CMB experiments improve, these probes become increasingly powerful and relevant to precision cosmology measurements. For example, the constraints on $\Omega_{\rm{m}}$ (the matter density parameter) and $\sigma_8$ (the amplitude of the linear power spectrum on the scale of $8h^{-1}$Mpc) from the ACT CMB lensing alone have achieved precision comparable to current weak lensing results (DES Y3 / KiDS-1000 / HSC-Y3) \citep[see Sect.~3 of][]{2023arXiv230405203M}. Besides the cosmological information carried by the CMB secondary effects, cross-correlating these signals with large-scale structure probes, such as galaxy positions and shapes, provides measurements with an independent set of systematic errors and improves the calibration of systematics in galaxy surveys. The role of CMB lensing cross-correlations has been emphasised in the context of SO \citep{2019JCAP...02..056A} and CMB-S4 \citep{2016arXiv161002743A}. There is a wealth of literature studying the synergy of CMB lensing and galaxy fields and the synergy of tSZ with galaxy fields. The theory of the former has recently been studied in the context of upcoming galaxy surveys and CMB experiments \citep[\eg,][]{2017PhRvD..95l3512S,2020JCAP...12..001S,2018PhRvD..97l3540S, 2022MNRAS.509.5721F,2022MNRAS.512.5311W}, and the synergies have been carried out in experiments, including SDSS + \planck \citep[\eg,][]{2017MNRAS.464.2120S}, DES + \planck \citep[\eg,][]{2016MNRAS.456.3213G,2019PhRvD.100d3501O,2019PhRvD.100d3517O,2019PhRvD.100b3541A}, DES + \planck + SPT \citep[\eg][]{2023PhRvD.107b3529O,2023PhRvD.107b3530C,2023PhRvD.107b3531A}, unWISE + \planck \citep[\eg,][]{2021JCAP...12..028K}, HSC + \planck \citep[\eg,][]{2022PhRvL.129f1301M}, KiDS + ACT + \planck \citep[\eg,][]{2021A&A...649A.146R}, DESI / DESI-like + \planck \citep[\eg,][]{2021MNRAS.501.1481H,2021MNRAS.501.6181K,2022JCAP...02..007W}. The synergies of tSZ with galaxy fields have also been explored both in theory \citep[\eg,][]{2020PhRvD.101d3525P,2022JCAP...04..046N} and in observations, such as Canada France Hawaii Lensing Survey + \planck \citep{2015JCAP...09..046M}, 2MASS + \planck \citep{2018MNRAS.480.3928M}, 2MASS + WISE\x SuperCOSMOS + \planck \citep{2020MNRAS.491.5464K}, SDSS + \planck \citep[\eg,][]{2017MNRAS.467.2315V,2018PhRvD..97h3501H,2020MNRAS.491.2318T,2020ApJ...902...56C}, SDSS + ACT \citep[\eg,][]{2021PhRvD.103f3513S}, KiDS + \planck + ACT \citep[\eg,][]{2021A&A...651A..76Y,2022A&A...660A..27T}, and DES + \planck + ACT \citep[\eg,][]{2022PhRvD.105l3525G,2022PhRvD.105l3526P}. However, there has not been any fully joint analysis combining all these probes, and simultaneously constraining cosmological and baryonic properties with a consistent halo model.

This paper explores the synergies of a CMB\x LSS joint analysis using the Simons Observatory (SO) and Vera C. Rubin Observatory's Legacy Survey of Space and Time (LSST) as examples. We perform the first ``10\x2pt'' simulated analysis, combining two-point functions between fields of galaxy density, shear, CMB lensing convergence, and tSZ Compton-$y$, which extends our previous work on 6\x2pt analyses combining galaxy density, shear, CMB lensing convergence \citep{2022MNRAS.509.5721F}.\footnote{We note that the term $N$\x2pt analysis is ambiguous, as $N=3/6/10$ different two-point correlations can be constructed from any two/three/four different tracer fields; e.g., other 6\x2pt combinations are possible, such as replacing CMB lensing with cluster density \citep{2021PhRvL.126n1301T}.} We present our theoretical modelling of the combined ``10\x2pt'' probes in Sect.~\ref{sec:multiprobe}, including the analysis choices, data vector, and analytic covariance modelling. We carry out the simulated likelihood forecast analysis for LSST + SO in Sect.~\ref{sec:results}. We compare the cosmological constraining power of 10\x2pt with 3\x2pt and 6\x2pt and identify the subset of probes containing most cosmological information. We also identify a subset of halo parameters responsible for improving the self-calibration between probes related to matter distribution (\ie, lensing and galaxy clustering) and probes related to gas distribution (\ie, tSZ). We conclude in Sect.~\ref{sec:discussion}.

In a companion paper (Fang et al. in prep, Paper II), we further investigate the 10\x2pt synergies between various scenarios of the Roman Space Telescope galaxy survey and SO/S4, including the benefits of choosing an alternative galaxy sample and the impact of halo model uncertainties.

\section{Multi-Survey Multi-Probe Analysis Framework}\label{sec:multiprobe}
This section describes the ingredients for simulated 10\x2pt analyses with focus on the tSZ auto- and cross-correlation modelling, building on the 3\x2pt and 6\x2pt analyses modelling and inference described in \cite{2017MNRAS.470.2100K} and \cite{2022MNRAS.509.5721F}.

\subsection{Survey Specifications}\label{ssec:choice}
\paragraph*{Simons Observatory}
The Simons Observatory (SO, \citealp{2018SPIE10708E..04G,2019JCAP...02..056A}) is a CMB experiment under construction in the Atacama desert in Chile, at an altitude of 5,200~m.
It is designed to observe the microwave sky in six frequency bands centred around 30, 40, 90, 150, 230, and 290~GHz, in order to separate the CMB from Galactic and extragalactic foregrounds.

The nominal design of the observatory will include one 6~m large-aperture telescope (LAT, \citealp{2021RNAAS...5..100X,2018SPIE10700E..41P}) and three small-aperture 0.5~m telescopes (SATs, \citealp{2020JLTP..200..461A}).
The LAT will produce temperature and polarisation maps of the CMB with $\sim$arcmin resolution over 40\% of the sky, with a sensitivity of $\sim$6 $\mu$K$\cdot$arcmin when combining 90 and 150~GHz bands \citep{2021RNAAS...5..100X}. 
These wide-field maps will be the key input to measure CMB lensing with SO. We also note that a large investment was recently announced that will significantly increase the detector count of the LAT and double the number of SATs, an upgrade known as ``Advanced Simons Observatory'' (ASO). Since little technical information is publicly available about ASO at present, our forecasts adopt the nominal configuration, and are therefore conservative. Further gains can be expected with ASO. We present the details of the noise levels and component separation techniques used in our analysis in App.~\ref{ssec:cov}.

\paragraph*{Vera C. Rubin Observatory's Legacy Survey of Space and Time }  
will rapidly and repeatedly cover a $\sim$ 18,000 deg$^2$ footprint in six optical bands (320 nm-1050 nm). With a single exposure depth of 24.7 r-band magnitude (5$\sigma$ point source), the 10 years of operations will achieve an overall depth of 27.5 r-band magnitude. 

The performance of the LSST dark energy analysis given specific analysis choices is explored in the LSST-Dark Energy Science Collaboration's (LSST-DESC) Science Requirements Document \citep[DESC-SRD,][]{2018arXiv180901669T}.

For galaxy clustering and weak lensing analyses, we adopt the galaxy samples and the survey parameters from \cite{2022MNRAS.509.5721F}, which is largely based on the DESC-SRD. We assume that LSST Y6 will cover the same area as the final SO Y5, \ie, 40\% of the sky or $\Omega_{\rm s}=16,500\,$deg$^2$. We also assume an $i$-band depth $i_{\rm depth}=26.1\,$mag for the weak lensing analysis, and $i$-band limiting magnitude $i_{\rm lim}=i_{\rm depth}-1= 25.1\,$mag for the clustering analysis.

\begin{figure}
    \centering
    \includegraphics[width=3.25in]{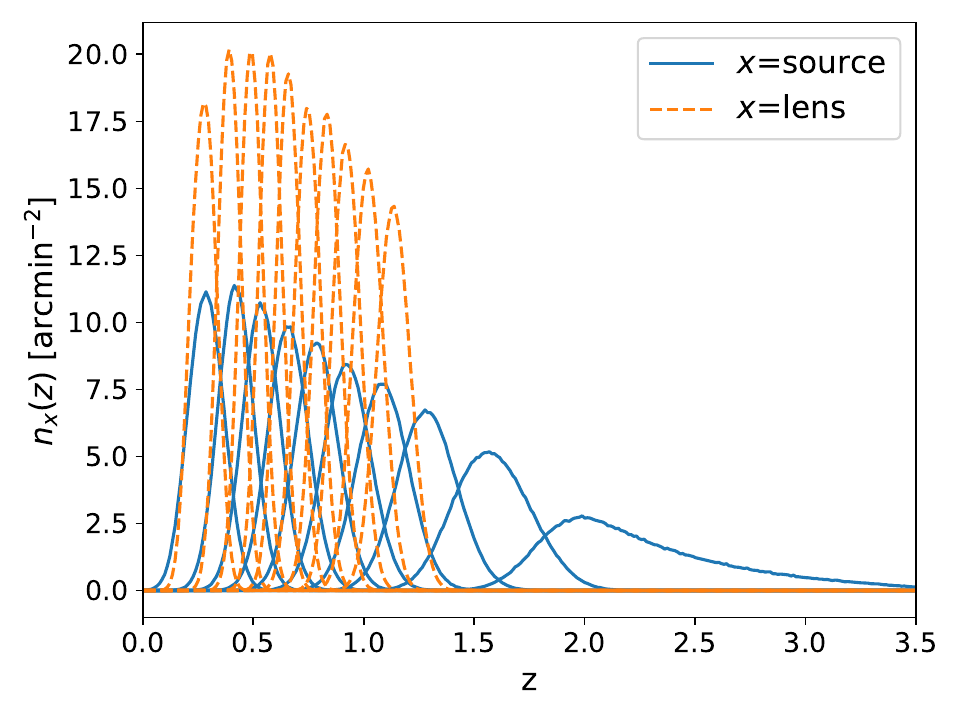}
    \caption{\label{fig:nz}The equally binned true redshift distributions $n_x(z)=\frac{dN_x}{dz d\Omega}$ of the source and lens samples of LSST Year 6, normalised by the corresponding binned effective number density $\bar{n}_x^{i}$ after redshift cuts. A fiducial Gaussian photo-$z$ error has been convolved with each tomographic bin as described by Eq.~(\ref{eq:binned-nz-true}).}
\end{figure}

We parameterise the photometric redshift distributions of both the lens and source samples as
\begin{equation}
    n_x(z_{\rm ph}) \equiv \frac{dN_x}{dz_{\rm ph}d\Omega} \propto z_{\rm ph}^2\exp[-(z_{\rm ph}/z_0)^\alpha]~,~~x\in\{{\rm lens,~source}\}
    \label{eq:nz-true}
\end{equation}
normalised by the effective number density $\bar{n}_x$. $N_x$ is the number counts of lens/source galaxies, $z_{\rm ph}$ is the photometric redshift, $\Omega$ is the solid angle. The parameter values of $\bar{n}_x,~z_0,~\alpha$ are given in Tab.~\ref{tab:nz}, which are updated values from the LSST-DESC's Observing Strategy\footnote{\url{https://github.com/LSSTDESC/ObsStrat/tree/static/static}} \citep{2018arXiv181200515L}. 
These values, slightly different from the DESC-SRD, are computed for the more recent optimisation studies of LSST observing strategies \citep{2018arXiv181200515L}. Following the redshift cuts in DESC-SRD ($z_{\rm min}=0.2$, $z_{\rm max}=1.2$ for the lens sample, $z_{\rm min}=0.2$ for the source sample), we further divide each galaxy sample into $N_{\mathrm{tomo}}=10$ equally populated tomographic bins as shown in Fig.~\ref{fig:nz}.

\begin{table}
    \centering
    \caption{\label{tab:nz}Parameters for generating the lens and source galaxy samples of LSST Y6.}
    \begin{tabular}{c c c}
    \hline\hline
    Parameter  & $x$=lens & $x$=source \\
         \hline
    $\bar{n}_{x}$ (arcmin$^{-2}$)  & 41.1 & 23.2 \\
    $z_0$ & 0.274 & 0.178\\
    $\alpha$ & 0.907 & 0.798\\
    \hline\hline
    \end{tabular}
\end{table}

\subsection{Multi-probe modelling}
We extend the existing 6\x2pt model of the {\py CosmoLike}\footnote{\url{https://github.com/CosmoLike}} modelling and inference framework \citep{2014MNRAS.440.1379E,2017MNRAS.470.2100K} to include tSZ (cross-) correlations. To obtain an (internally) consistent model of all 10\x2pt statistics, we adopt a halo model \citep[][c.f. App.~\ref{sssec:halo-model}]{2000MNRAS.318..203S,Ma_HM,Peacock_HM,2002PhR...372....1C, Hill:2013baa, Hill:2013dxa} approach and specify all systematics at the level of the individual observables, \ie, the density contrast of lens galaxies $\delta_{\rm g}$, the lensing convergence of source galaxies $\kappa_{\rm g}$, the CMB lensing convergence $\kappa_{\rm CMB}$, and  the Compton-$y$ field $y$. We summarise the well-established computation of angular power spectra and our model ingredients for $\delta_{\rm g}$, $\kappa_{\rm g}$ and $\kappa_{\rm CMB}$, including astrophysical systematics linear galaxy bias model, intrinsic alignments (IA) using the ``nonlinear linear alignment'' model \citep{2004PhRvD..70f3526H,2007NJPh....9..444B},  and observational systematics, in App.~\ref{ssec:2pt}. Combined with tSZ and baryonic feedback modelling described in this subsection, our 10\x2pt model has 57 free parameters, which are listed in Tab.~\ref{tab:params}.

\begin{table}
\footnotesize
    \centering
    \caption{\label{tab:params}A list of the parameters characterising the surveys, cosmology and systematics. The fiducial values are used for generating the simulated data vectors, and the priors are used in the sampling. Uniform priors are described by $U$[minimum, maximum], and Gaussian priors are described through the normal distribution $\mathcal{N}(\mu, \sigma^2)$.}
    \begin{tabular}{l l l }
    \hline\hline
    Parameters & Fiducial & Prior \\
    \hline
    \multicolumn{3}{l}{\textbf{Survey}} \\
    $\Omega_{\rm s}$ (deg$^2$) & 16500 & fixed \\
    $\sigma_e$ & 0.26/component & fixed \\
    \hline
    \multicolumn{2}{l}{\textbf{Cosmology}} & \\
    $\Omega_{\rm m}$ & 0.3156 & $U$[0.05, 0.6]\\
    $\sigma_8$ & 0.831 & $U$[0.5, 1.1] \\
    $n_{\rm s}$ & 0.9645 & $U$[0.85, 1.05]\\
    $\Omega_b$ & 0.0492 & $U$[0.04, 0.055]\\
    $h_0$ & 0.6727 & $U$[0.4, 0.9] \\
    \hline
    \multicolumn{2}{l}{\textbf{Galaxy Bias}} & \\
    $b^i$ ${\scriptstyle(i=1,\cdots,N_{\mathrm{tomo}})}$ & $0.95/G(\bar{z}^i)$ & $U$[0.4, 3] \\
    \hline
    \multicolumn{2}{l}{\textbf{Photo-$z$}} & \\
    $\Delta_{z,\rm lens}^i$ ${\scriptstyle(i=1,\cdots,N_{\mathrm{tomo}})}$ & 0 & $\mathcal{N}(0, 0.001^2)$ \\
    $\sigma_{z,\rm lens}$ & 0.03 & $\mathcal{N}(0.03, 0.003^2)$ \\
    $\Delta_{z,\rm source}^i$ ${\scriptstyle(i=1,\cdots,N_{\mathrm{tomo}})}$ & 0 & $\mathcal{N}(0, 0.001^2)$ \\
    $\sigma_{z,\rm source}$ & 0.05 & $\mathcal{N}(0.05, 0.003^2)$ \\
    \hline
    \multicolumn{2}{l}{\textbf{Shear Calibration}} & \\
    $m^i$ ${\scriptstyle(i=1,\cdots,N_{\mathrm{tomo}})}$ & 0 & $\mathcal{N}(0, 0.003^2)$ \\
    \hline
    \multicolumn{2}{l}{\textbf{IA}} &  \\
    $A_{\rm IA}$ & 0.5 & $U$[-5, 5]\\
    $\eta_{\rm IA}$ & 0 & $U$[-5, 5]\\
        \hline
    \multicolumn{2}{l}{\textbf{Halo and Gas Parameters}} & \\
    $\Gamma$ & 1.17 & $U$[1.05, 1.35]\\
    $\beta$ & 0.6 & $U$[0.2, 1.0] \\
    $\log_{10}M_0$ & 14.0 & $U$[12.5, 15.0]\\
    $\alpha$ & 1.0 & $U$[0.5, 1.5]\\
    $\log_{10} T_w$ & 6.5 & $U$[6.0, 7.0] \\
    $\epsilon_1$ & 0 & $U$[-0.8, 0.8]\\
    $\epsilon_2$ & 0 & $U$[-0.8, 0.8]\\
    $f_{\rm H}$ & 0.752 & $U$[0.7, 0.8] \\
    $M_{\rm min}, M_{\rm max}$ & $10^6, 10^{17}M_\odot/h$ & fixed \\
    $A_*$ & 0.03 & fixed \\
    $\log_{10}M_*$ & 12.5 & fixed \\
    $\sigma_*$ & 1.2 & fixed \\
    \hline\hline
    \end{tabular}
\end{table}
\subsubsection{Modelling baryons and tSZ cross-correlations}
Our implementation of tSZ cross-correlations and baryonic feedback effects on the matter distribution follows closely the empirical HMx halo model parameterisation for matter and pressure \citep{2020A&A...641A.130M}. The parameters of the HMx model are calibrated so as to ensure agreement at the power spectrum level between the prediction of the HMx prescription and the BAHAMAS simulations \citep{2017MNRAS.465.2936M}, which are a set of smooth-particle hydrodynamical simulations with separate dark matter and gas particles, including sub-grid recipes describing effects due to stars, supernovae, AGNs, and gas heating and cooling. As the dominant contribution to pressure (cross-) power spectra comes from massive clusters, an accurate calibration of the pressure power spectrum would require very large simulation volumes. The accuracy of HMx prescription for the matter-pressure cross-spectrum is of order $\sim15\%$ and worse for the pressure auto-power spectrum \citep{2020A&A...641A.130M}. While Stage-IV data analyses will certainly require refinements of the model parameterisation and parameter calibration, we use the HMx parameterisation only as a toy model to qualitatively explore information content in the non-linear regime as well as the impact of parameter self-calibration on cosmology constraints.

The halo model implementation in {\py CosmoLike} uses the \citet[NFW]{1997ApJ...490..493N} profile and the \cite{2008MNRAS.390L..64D} halo mass-concentration relation to model the halo density profile $\tilde{u}_\delta(k,M,z)$, the \cite{2010ApJ...724..878T} fitting functions to compute the linear halo bias $b_{h,1}(M)$ and the halo mass function $dn(M)/dM$. 

We use the $\Delta_{\bar{\rho}}=200$ halo definition, so that the halo radius $r_{200}$ satisfies $M=4\pi r_{200}^3\bar{\rho}\Delta_{\bar{\rho}}/3$ and the halo scale radius is given by $r_s=r_{200}/c(M,z)$. We note that several of these halo model ingredients differ from those assumed by \cite{2020A&A...641A.130M}. However, we do not expect these differences to qualitatively change the parameter degeneracy structure relevant for self-calibration studies presented here. Our main goal is indeed to investigate, within a coherent modelling choice, the interplay between the different model parameters and estimate how much the inter-calibration between datasets improves the constraints of the parameters of interest. We expect that, even with its limitations, our model is representative enough to allow this study.

\paragraph*{Electron pressure}
The Fourier-transformed $y$ halo profile, $\tilde{u}_y(k,M,z)$, which is the building block for halo model calculations of tSZ (cross-) power spectra (c.f.~App.~\ref{sssec:halo-model}) is then computed from the electron pressure profile $\mathcal{P}_{\rm e}(M,r,z)$,
\begin{equation}
    \tilde{u}_y(k,M,z)=4\pi \int_0^\infty dr\,r^2j_0(kr)\mathcal{P}_{\rm e}(M,r,z)~.
\end{equation}

The HMx halo model parameterisation describes the gas distribution with two components, bound and ejected. The bound gas is modelled with the Komatsu-Seljak (KS) profile \citep{2002MNRAS.336.1256K}, while the ejected gas is assumed to follow the linear perturbations of the matter field. Therefore, the ejected gas only contributes to the 2-halo term.

The fractions of bound and ejected gas in halos are parameterised as
\begin{align}
    f_{\rm bnd}(M)&=\frac{\Omega_{\rm b}}{\Omega_{\rm m}}\frac{1}{1+(M_0/M)^\beta}~,\label{eq:fbnd}\\
    f_{\rm ejc}(M)&=\frac{\Omega_{\rm b}}{\Omega_{\rm m}}-f_{\rm bnd}(M)-f_*(M)\nonumber\\
    &=\frac{\Omega_{\rm b}}{\Omega_{\rm m}}\frac{1}{1+(M_0/M)^{-\beta}}-A_* \exp\left[-\frac{\log_{10}^2(M/M_*)}{2\sigma_*^2}\right]~,\label{eq:fejc}
\end{align}
where $M_0$ and $\beta$ are an empirical parameterisation of the dependence of the fraction of ejected gas on halo mass, and the stellar fraction $f_*(M)$ is assumed to take a log-normal form, parameterised by $A_*, M_*, \sigma_*$). $\Omega_{\rm b}$ and $\Omega_{\rm m}$ are the cosmic baryon and matter density parameters. 

The bound gas density profile $\rho_{\rm bnd}$ can then be written as
\begin{equation}
    \rho_{\rm bnd}(M,r,z) = A(M,z) Mf_{\rm bnd}(M)\left[\frac{\ln(1+r/r_s)}{r/r_s}\right]^{1/(\Gamma-1)}~,
\end{equation}
where the radial dependence follows the KS profile with polytropic index $\Gamma$ \citep{2002MNRAS.336.1256K}. The normalisation is given $\int_0^{r_{\rm v}}dr\,4\pi r^2\rho_{\rm bnd}(M,r,z)=Mf_{\rm bnd}(M)~$, leading to
\begin{equation}
    [A(M,z)]^{-1} = \int_0^{r_{\rm v}}dr\,4\pi r^2\left[\frac{\ln(1+r/r_s)}{r/r_s}\right]^{1/(\Gamma-1)}~.
\end{equation}
In the KS profile the bound gas temperature $T_{\rm g}$ is determined by hydrostatic equilibrium,
\begin{equation}
T_{\rm{g}}(M,r,z) =T_{\rm v}(M,z)\frac{\ln (1+r/r_s)}{r/r_s}\,
\end{equation}
with $T_{\rm v}(M,z)$ the halo virial temperature
\begin{equation}
    \frac{3}{2}k_B T_{\rm v}(M,z)= \alpha\frac{GM m_{\rm p}\mu_{\rm p}}{ r_{\rm v}}(1+z)~,
\end{equation}
where $m_{\rm p}$ is the proton mass, $\mu_{\rm p} = 4/(3 + 5 f_{\rm H})$, with $f_{\rm H}$ the hydrogen mass fraction, and $\alpha$ is a free parameter that encapsulates the deviations from the hydrostatic equilibrium due to non-thermal components of the gas. The electron pressure profile of the bound gas can then be written as
\begin{align}
    \mathcal{P}_{\rm e,bnd}(M,r,z)&=\frac{\rho_{\rm bnd}(M,r,z)}{m_{\rm p}\mu_e}k_B T_{\rm g}(M,r,z)\nonumber\\
    &=\frac{\rho_{\rm bnd}(M,r,z)}{m_{\rm p}\mu_e}k_B T_{\rm v}(M,z)\frac{\ln (1+r/r_s)}{r/r_s}~,
\end{align}
where $\mu_e=2/(1+f_{\rm H})$, and $k_B$ is the Boltzmann constant.
The density profile for the ejected gas $\rho_{\rm ejc}$ in 2-halo term is approximated as a 3D Dirac delta function distribution with a constant temperature $T_{\rm w}$ and total mass $M_{\rm ejc}$, following the footnote 7 of \cite{2020A&A...641A.130M},
\begin{align}
    \mathcal{P}_{\rm e,ejc}(M,\bv{r},z)&=\frac{\rho_{\rm ejc}(M,r,z)}{m_{\rm p}\mu_e}k_B T_{\rm w}=\frac{M_{\rm ejc}(M,z)\delta_{\rm D}(\bv{r})}{m_{\rm p}\mu_e}k_B T_{\rm w}~.
\end{align}

\paragraph*{Impact of baryonic feedback on the matter distribution}
To account for the impact of baryonic feedback on the dark matter distribution, we follow the HMx approach which modifies the dark matter-only halo mass-concentration ($c_{\scriptscriptstyle\rm D}(M,z)$) 
\begin{align}
    c(M,z) &=c_{\scriptscriptstyle\rm D}(M,z)\left[1+\epsilon_1+(\epsilon_2-\epsilon_1)\frac{f_{\rm bnd}(M)}{\Omega_{\rm b}/\Omega_{\rm{m}}}\right]\nonumber\\
    &=c_{\scriptscriptstyle\rm D}(M,z)\left[1+\epsilon_1+\frac{\epsilon_2-\epsilon_1}{1+(M_0/M)^\beta}\right]~,
\label{eq:cM}
\end{align}
 where $\epsilon_1,\epsilon_2$ are free parameters and $\epsilon_1=\epsilon_2=0$ reduces to the unmodified case.

\paragraph*{Stellar component}
Throughout the analyses presented here, the HMx halo model parameters $A_*, M_*, \sigma_*$, which describe the stellar mass fraction, are held fixed as the 10\x2pt data vector has almost no sensitivity to these parameters.

\subsection{Simulated likelihood analysis methodology}
To simulate joint analyses of multi-survey multi-probe data, we perform simulated parameter inference assuming a Gaussian likelihood for the data $\bv{d}$ given a point $\bv{p}$ in cosmological and nuisance parameter space,
\begin{equation}
    L(\bv{d}|\bv{p}) \propto \ee^{-\chi^2/2}~,~~{\rm where}~~\chi^2=[\bv{d}-\bv{m}(\bv{p})]^\top \bmat{C}^{-1}[\bv{d}-\bv{m}(\bv{p})]~,
\end{equation}
where $\bv{m}$ is the model vector and $\bmat{C}$ is the data covariance matrix. The synthetic (noiseless) data is calculated from the input model, with cosmic variance and shape/shot noise entering only through the covariance; this setup bypasses scatter in the inferred parameters due to realisation noise and allows us to focus on the constraining power of different analysis choices. The analytic Fourier space 10\x2pt covariance calculation is summarised in App.~\ref{ssec:cov} and we make available an implementation {\py CosmoCov\_Fourier}\footnote{\url{https://github.com/CosmoLike/CosmoCov_Fourier}}, which is an extension of {\py CosmoCov}\footnote{\url{https://github.com/CosmoLike/CosmoCov}} \citep{2020MNRAS.497.2699F}. 
\paragraph*{Scale cuts} We compute all two-point functions in 25 logarithmically spaced Fourier mode bins ranging from $\ell_{\rm min}=20$ to $\ell_{\rm max}=8000$, and impose the following set of $\ell$-cuts:
\begin{itemize}
    \item \textbf{Galaxy clustering $\delta_{\rm g}$} scale cuts are driven by the modelling inaccuracy of non-linear galaxy biasing. Following the DESC-SRD, we adopt $\ell^i_{\rm max}=k_{\rm max}\,\chi(\bar{z}^i)$, where $k_{\rm max}=0.3\,h/$Mpc and $\bar{z}^i$ is the mean redshift of the lens bin $i$.
    \item \textbf{Weak lensing $\kappa_{\rm g}$} scale cuts are driven by model misspecifications for the impact of baryonic processes on the non-linear matter power spectrum as well as gravitational non-linearity.
    
    We assume $\ell_{\rm max}=3000$ for all tomography bins, following the DESC-SRD. We also consider a more optimistic case in which we can reliably model shear up to $\ell_{\rm max}=8000$. We note that baryonic effects on the shear power spectrum will be significant for both of these scale cuts, and Stage-IV analyses will likely require modelling beyond the modified halo mass-concentration relation (Eq.~\ref{eq:cM}) implemented here to reach these scale cuts, these different $\ell_{\rm max}$ choices will illustrate the interplay of scale cuts and parameter degeneracies. 
    \item \textbf{CMB lensing $\kappa_{\rm CMB}$} measurements are challenging at high $\ell$ due to lensing reconstruction challenges from foregrounds; similarly, various systematic errors can limit the reconstruction of low-$\ell$ lensing modes \citep{2021MNRAS.500.2250D}. We adopt the scale cuts $\ell_{\rm min}=100$, $\ell_{\rm max}=3000$.\footnote{While the interpretation of CMB lensing (cross-) spectra is of course also limited by the same high-$\ell$ processes that set the galaxy weak lensing scale cuts, the fractional impact of these processes on the power spectrum decreases with redshift. Hence $\ell_{\rm max}=3000$ for $\kappa_{\rm g}$ would correspond to a less restrictive scale cut for $\kappa_{\rm CMB}$.} More details on the lensing reconstruction are presented in App.~\ref{ssec:cov}.
    \item \textbf{tSZ} $y$ measurements are limited by atmospheric noise at low-$\ell$ (in the case of SO) and component separation at high-$\ell$ \citep{2019JCAP...02..056A}. We adopt the somewhat optimistic scale cuts $\ell_{\rm min}=80$, $\ell_{\rm max}=8000$, noting in particular that contamination from the Cosmic Infrared Background (CIB) will need to be modelled and (likely) marginalised over.
\end{itemize}
For cross-power spectra, we adopt the more restrictive scale cut combination of the two fields. We further exclude $\delta_{\rm g}$-$\kappa_{\rm g}$ combinations without cosmological signal, \ie, with the lens tomography bin at higher redshift than the source tomography bin.

\subsubsection{Analytical Marginalisation of photo-$z$ and Shear Calibration Parameters}\label{subsec:analytic-margin}
To speed up our forecast analyses, we analytically marginalise the 22 photo-$z$ parameters and 10 shear calibration parameters, assuming that constraints on these parameters are dominated by their (Gaussian) priors listed in Tab.~\ref{tab:params}. This analytic marginalisation was derived in the context of cosmological surveys in \cite{2002MNRAS.335.1193B}, which we summarise below \citep[see also][]{IMM2012-03274}.

Suppose that the likelihood $L$ of obtaining the $N$-dimensional data vector $\bv{d}$ given the model vector $\bv{m}(\bv{p},\bm{\lambda})$ is a multivariate Gaussian with covariance $\bmat{C}$, \ie, $\bv{d}|\bv{m}(\bv{p},\bm{\lambda})\sim \mathcal{N}(\bv{m}(\bv{p},\bm{\lambda}), \bmat{C})$, where $\bv{p},\bm{\lambda}$ are parameters of the model, and $\bm{\lambda}$ contains the parameters we want to marginalise over, with a joint prior distribution $p(\bm{\lambda})$. Marginalising the posterior distribution over $\bm{\lambda}$ has a simple analytic solution if we assume (1) the model is linear, and (2) the prior $p(\bm{\lambda})$ is a multivariate Gaussian distribution.

Assuming that $p(\bm{\lambda})$ follows $\mathcal{N}(\bm{\lambda}_0,\bmat{C}_{\bm{\lambda}})$, and a linear model
\begin{equation}
    \bv{m}(\bv{p},\bm{\lambda}) = \bv{m}(\bv{p},\bm{\lambda}_0) + \bmat{J}\bdot (\bm{\lambda}-\bm{\lambda}_0)~,
\end{equation}
where the Jacobian matrix $\bmat{J}$ is defined by $\bmat{J}=\bm{\nabla}_{\bm{\lambda}}\bv{m}(\bv{p},\bm{\lambda})|_{\bm{\lambda}=\bm{\lambda}_0}$. The marginalised posterior can be derived as $\bv{d}|\bv{m}(\bv{p},\bm{\lambda}_0)\sim \mathcal{N}(\bv{m}(\bv{p},\bm{\lambda}_0), \bmat{C}_{\rm mod})$, where
\begin{equation}
    \bmat{C}_{\rm mod} = \bmat{C} + \bmat{J}\,\bmat{C}_{\bm{\lambda}}\,\bmat{J}^\btop~.
\end{equation}
Therefore, we can eliminate the parameters $\bm{\lambda}$ and replace the covariance $\bmat{C}$ with the modified version $\bmat{C}_{\rm mod}$ in the likelihood.

For a non-linear model, the linear model is still a good approximation when the prior of $\bm{\lambda}$ is narrow around $\bm{\lambda}_0$, such that higher-order expansion can be neglected. We numerically compute the Jacobian matrix $\bmat{J}$ for the 32 nuisance parameters around their fiducial values.

This method reduces the number of sampled parameters from 57 to 25, enabling quick convergence at the cost of our ability to investigate the self-calibration of those 32 nuisance parameters; we apply this method throughout this paper.

\section{Simulated Likelihood Analysis Results}\label{sec:results}
For each shear scale cut choice, $\ell_{\rm max}^{\rm shear}=500$ or 3000 or 8000, we run simulated likelihood analyses of 3\x2pt, 6\x2pt, 10\x2pt, and 8\x2pt, where the latter analysis drops the $\kappa_{\rm CMB}$-$y$ and $\delta_{\rm g}$-$y$ combinations compared to 10\x2pt analysis that includes all possible cross-correlations. We use {\py emcee} \citep{2013PASP..125..306F} to sample the parameter space.

In this section, we present a series of simulated analyses that quantify and interpret the gain in constraining unlocked in 10\x2pt analyses: We first quantify the measurement signal-to-noise ratio and cosmological constraining power in the context of our baseline halo model for different probe combination. We then design simulated analyses that explore the degradation of cosmology parameter constraints due to halo model parameter uncertainties, which limit the cosmological interpretation of small-scale measurements, and isolate the importance of baryonic feedback parameter self-calibration.
\subsection{Constraining power of multi-probe combinations}\label{ssec:constraints}
\begin{figure}
    \centering
    \includegraphics[width=3.25in]{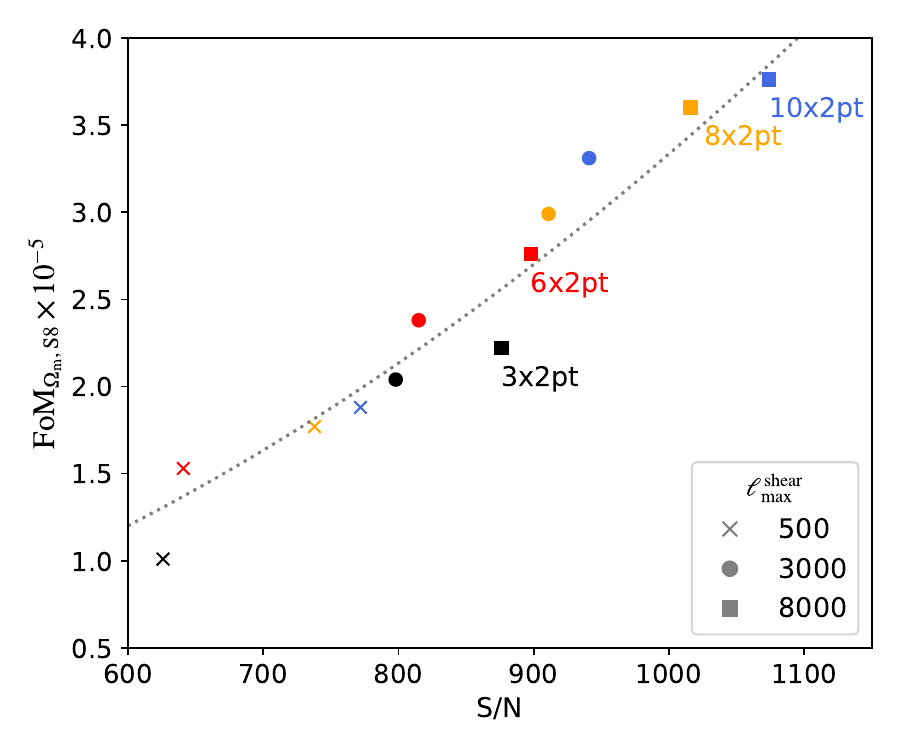}
\caption{The signal-to-noise ratio (S/N) and $\Omega_{\rm{m}}-S_8$ Figure-of-Merit (FoM) of different combined probes for 3 shear scale cut choices $\ell_{\rm max}^{\rm shear}=500$, 3000 and 8000. The dotted line shows the $\rm{FoM}\propto(\rm{S/N})^2$ scaling.}
    \label{fig:sn_fom}
\end{figure}

The signal-to-noise ratios (S/N), defined by S/N $=\sqrt{\bv{d}\,\bmat{C}^{-1}\,\bv{d}^\btop}$, of different measurements can be used as a simple proxy for their constraining power. For a single-parameter model, S/N is directly related to the constraining power on this parameter; for complex models, degeneracies between parameters may degrade the constraining power on parameters compared to simple S/N expectations.
We evaluate the S/N with the data computed at the fiducial parameter values and show the results for different multi-probe combination with different $\ell_{\rm max}^{\rm shear}$ in Fig.~\ref{fig:sn_fom}.
With the scale cuts described in Sect.~\ref{ssec:choice}, we find the S/N to improve by $\sim 10-15\%$ when very small-scale shear information is included ($\ell_{\rm max}^{\rm shear}=8000$ vs. $\ell_{\rm max}^{\rm shear}=3000$), with only minor variations between different probe combinations. The addition of all CMB lensing and tSZ (cross-)correlations increases the S/N by $\sim 20\%$ compared to 3\x2pt data. Notably, there is limited information gain
from the $\kappa_{\rm CMB}$-$y$ and $\delta_{\rm g}$-$y$ cross-correlations, with the S/N increasing only $\sim 5\%$ from 8\x2pt to 10\x2pt.

\begin{figure}
    \centering
    \includegraphics[width=3.25in]{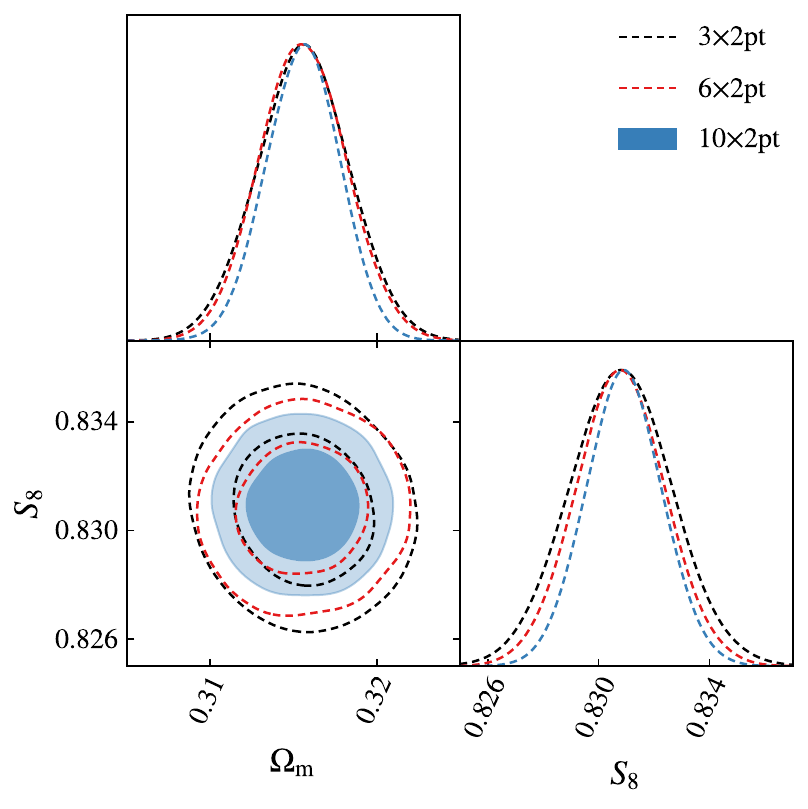}
    \caption{Increase in LSST Y6 + SO Y5 constraining power as the data vector increases from 3\x2pt (black dashed) to 6\x2pt (red dashed) to 10\x2pt (blue shaded), for galaxy shear scale cuts $\ell_{\rm max}^{\rm shear}=3000$.}
    \label{fig:om-s8_3_6_10}
\end{figure}

To illustrate the cosmological constraining power of different probe combinations, we focus on $\Omega_{\rm{m}}$ and $S_8 \equiv \sigma_8(\Omega_{\rm{m}}/\Omega^{\rm fid}_{\rm{m}})^{0.4}$, where the power index 0.4 is chosen such that $S_8$ captures the well-measured combination, as parameters of interest. We use the Figure-of-Merit (FoM) in the $\Omega_{\rm{m}}-S_8$ subspace, ${\rm FoM}_{\Omega_{\rm{m}},S_8} = 1/\sqrt{\det\,{\rm Cov}(\Omega_{\rm{m}}, S_8)}$, where the covariances ${\rm Cov}(\Omega_{\rm{m}}, S_8)$ are computed using the last 500k samples of the MCMC chains, as a simple metric for constraining power. The cosmological constraints from 3\x2pt, 6\x2pt, and 10\x2pt analyses are shown in Fig.~\ref{fig:om-s8_3_6_10}, with the corresponding FoM values shown in Fig.~\ref{fig:sn_fom}. The constraints improve by $\sim 50\%$ (in FoM) as CMB lensing and tSZ are included in the analysis while the corresponding gain in S/N is only $\sim 20\%$, which illustrates the importance of parameter self-calibration and breaking of degeneracies for parametric constraints. Fig.~\ref{fig:sn_fom} also shows the limited gain in constraining power from more aggressive shear scale cuts, which is consistent with the corresponding S/N gains.

We also find negligible difference in the $\Omega_{\rm{m}}-S_8$ constraints between 8\x2pt and 10\x2pt probes, which suggests that $\kappa_{\rm CMB}$-$y$ and $\delta_{\rm g}$-$y$ combinations may be dropped with limited impact on cosmology constraints. However, these combinations may contribute to self-calibration of additional systematic parameters in extended models.

\subsection{Impact of halo model uncertainties}\label{ssec:halo-impact}
\begin{figure}
    \centering
    \includegraphics[width=3.25in]{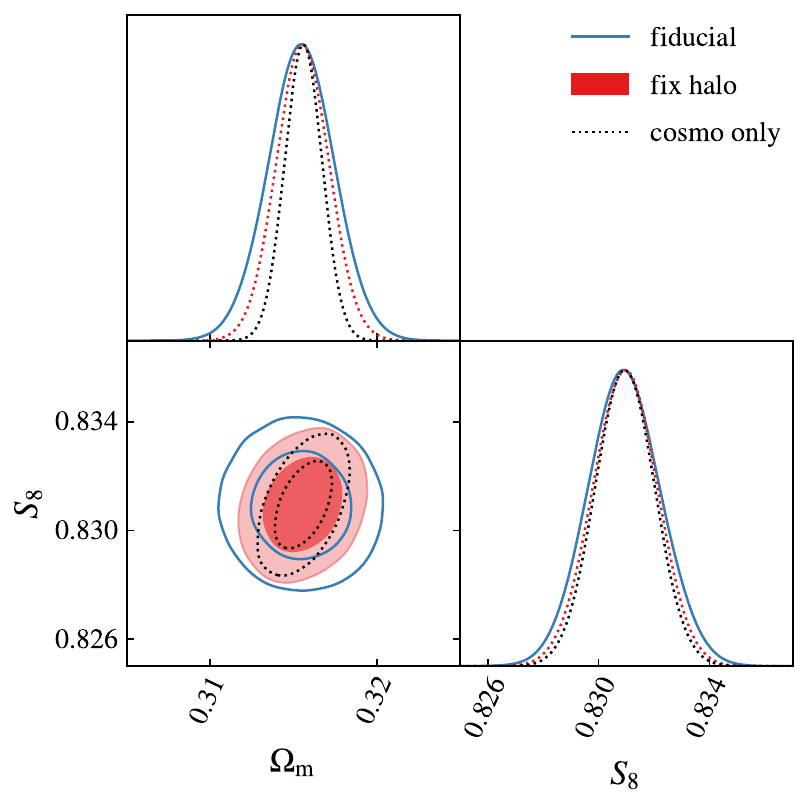}
    \caption{Degradation of the $\Omega_{\rm{m}}-S_8$ constraining power for 10\x2pt analyses with $\ell_{\rm max}^{\rm shear}=8000$ as a function of the model complexity. The ``fiducial'' model (blue line) samples over five cosmological parameters, 12 astrophysical nuisance parameters, and eight halo parameters (c.f. Tab.~\ref{tab:params}). The ``fix halo'' scenario (red shaded) fixes all halo parameters at their fiducial values, while the ``cosmo only'' scenario further fixes the remaining 12 nuisance parameters at their fiducial values.}
    \label{fig:fixhalo}
\end{figure}

To illustrate the degradation of cosmological constraints due to the systematic uncertainties, we run two additional simulated analyses: (1) ``fix halo'', fixing all the eight free halo parameters at their fiducial values, and (2)  ``cosmo only'', further fixing the IA and galaxy bias parameters and sampling only the five cosmological parameters. Note that in all the scenarios, the photo-$z$ and shear calibration parameters are analytically marginalised, which does not allow for internal calibration of these parameters and may thus underestimate the gains from decreasing model complexity. The $\Omega_{\rm{m}}-S_8$ constraints are shown in Fig.~\ref{fig:fixhalo} and the corresponding FoMs for ``fiducial'' and ``fix halo'' cases are $3.8\times10^{5}$ and $5.6\times10^{5}$, respectively.

One way to identify the halo model parameter(s) most responsible for degrading the cosmological constraints is to run chains similar to the “fix halo” analyses, whereby we fix one or more parameters and compute the impact on the FoM. Since this is a computationally expensive step, we instead utilise the full parameter covariance from the fiducial case.  We first approximate the posterior distribution of the fiducial case as a multivariate Gaussian with parameter covariance $\bmat{C}_{\bv{p}}$. Fixing a combination of one or more parameters $\tilde{\bv{p}}$, we analytically compute the  conditional covariance for the remaining parameters \citep{IMM2012-03274}. Suppose that the fixed parameters $\tilde{\bv{p}}$ correspond to the covariance block $\bmat{C}_{\tilde{\bm{p}}}$, then the remaining parameters’ joint posterior (\ie, the conditional distribution) will be a multivariate Gaussian with its covariance given by the Schur complement of block $\bmat{C}_{\tilde{\bm{p}}}$,\footnote{For a block matrix $\bmat{A}=\left(\begin{smallmatrix}
\bmat{A}_{11} & \bmat{A}_{12}\\ \bmat{A}_{21} & \bmat{A}_{22}\end{smallmatrix}\right)$, the Schur complement of block $\bmat{A}_{11}$ is given by $\bmat{A}_{22} - \bmat{A}_{21}\bmat{A}_{11}^{-1}\bmat{A}_{12}$.}  from which we can calculate the conditional FoMs. We find that in the limiting case when all eight halo parameters are fixed, the estimated FoM of $5.4\times10^{5}$ is consistent with the FoM measured directly from the ``fix halo'' chain.

We vary the number of fixed halo parameters $N_{\tilde{p}}$ and compute the conditional FoMs for all possible parameter combinations. We identify the combination that maximises the conditional FoM, which are shown in Tab.~\ref{tab:fom}. We find that $N_{\tilde{p}}=3$ is sufficient for the maximal conditional FoM to reach 90\% of the limiting FoM. These results identify that the parameters describing the halo mass-concentration relation ($\epsilon_1,\epsilon_2, M_0,\beta$, Eq.~\ref{eq:cM}) fractions of bound and ejected gas ($M_0,\beta$, Eqs.~\ref{eq:fbnd}-\ref{eq:fejc}) contribute the most to the degradation of constraining power on $\Omega_{\rm{m}}$ and $S_8$ parameters. Additionally, the concentration of the gas profile, parameterised by the gas polytropic index $\Gamma$, and non-thermal contributions to gas temperature, $\alpha$, can have a relatively high impact on constraining power as well. This test suggests that improved priors on the halo mass-concentration (from independent observations or/and hydrodynamic simulations) will have the highest impact on constraining power on cosmology.
\begin{table}
\footnotesize
    \centering
    \caption{\label{tab:fom} Estimated $\Omega_{\rm{m}}-S_8$ Figure-of-Merit (FoM) of 10\x2pt with galaxy shear scale cuts $\ell_{\rm max}^{\rm shear}=8000$ when the number $N_{\tilde{p}}$ of fixed halo parameters $\tilde{\bv{p}}$ increases.}
    \begin{tabular}{r c r}
    \hline\hline
     $N_{\tilde{p}}$ & fixed parameters $\tilde{\bv{p}}$ & ${\rm FoM}_{\Omega_{\rm{m}},S_8}/{\rm FoM}_{\rm limit}$ \\
     \hline
     0 & None & 70\% \\
     \hline
     1 & $\epsilon_1$ & 76\% \\
       & $\Gamma$ & 75\% \\
       & $x \in \{\alpha,\beta,\epsilon_2,f_{\rm{H}},M_{0} \}$ & 71\% \\
     \hline
     2 & $\epsilon_1,M_0$ & 84\% \\
       & $\epsilon_1,\beta$ & 84\% \\
       & $\epsilon_1,\epsilon_2$ & 82\% \\
     \hline
     3 & $\epsilon_1,\epsilon_2,\Gamma$ & 91\% \\
       & $\epsilon_1,\epsilon_2,\beta$ & 87\% \\
       & $\epsilon_1,M_0,\Gamma$ & 87\% \\
       & $\epsilon_1,\beta,\Gamma$ & 87\% \\
       & $\epsilon_1,\beta,\alpha$ & 87\% \\
       & $\epsilon_1,M_0,\alpha$ & 87\% \\
       & $\epsilon_1,\epsilon_2,M_0$ & 86\% \\
    \hline\hline
    \end{tabular}
\end{table}

\begin{figure}
    \centering
    \includegraphics[width=3.25in]{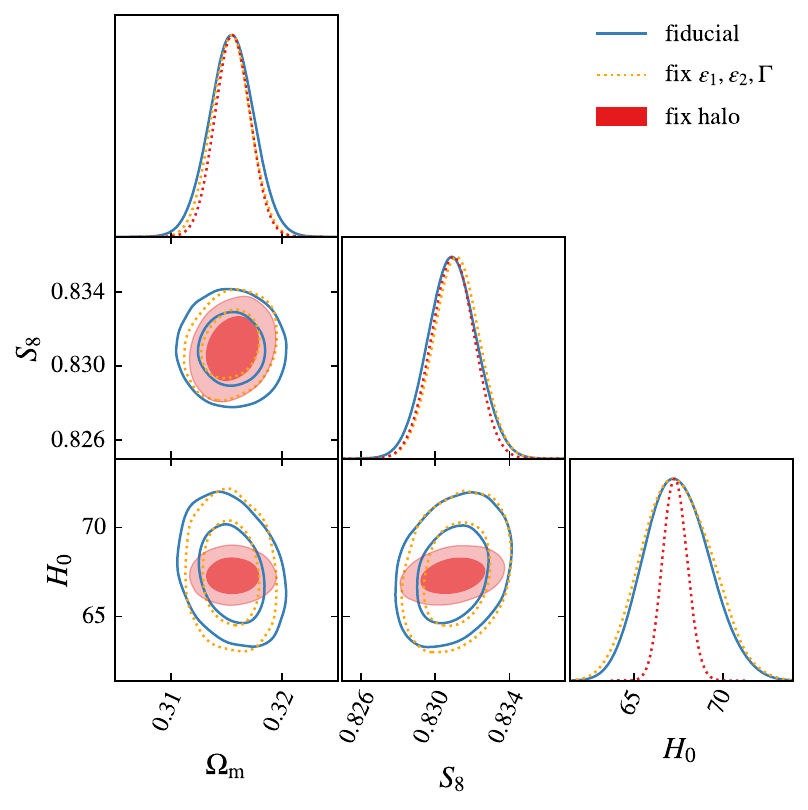}
    \caption{\label{fig:fixhalo_3param} Degradation of cosmological parameter constraints due to halo model uncertainties 10\x2pt analyses with $\ell_{\rm max}^{\rm shear}=8000$. In addition to ``fiducial'' model (blue line) and ``fix halo'' scenario (red shaded) repeated from Fig.~\ref{fig:fixhalo}, the orange dotted contour shows an intermediate model fixing only the three halo model parameters $\epsilon_1, \epsilon_2, \Gamma$. These three parameters most degrade the constraining power on $\Omega_{\rm{m}}$ and $S_8$ and are thus high-reward targets for external calibration.}
\end{figure}

As the posterior distribution of parameters is not a multivariate Gaussian, we show a simulated 10\x2pt analysis with the three highest-ranked halo parameters from Tab.~\ref{tab:fom} ($\epsilon_1, \epsilon_2, \Gamma$) fixed at their fiducial values in Fig.~\ref{fig:fixhalo_3param} (dashed orange line). The $\Omega_{\rm{m}}-S_8$ constraints approach the limiting case where all halo parameters are perfectly known (``fix halo'', solid red contour). As these halo parameters were selected by their impact on ${\rm FoM}_{\Omega_{\rm{m}},S_8}$, the rank ordering may be different for other parameters of interest, and our example performs poorly in recovering $H_0$ constraining power.

\subsection{Self-calibration of small-scale modelling}\label{ssec:calib}
Within the HMx-like halo model parameterization adopted here, the cosmological information gain from small scales is limited by (halo) modelling uncertainties, as demonstrated in Sect.~\ref{ssec:halo-impact}. When including tSZ information, the halo model parameters describing the matter field are self-calibrated from $y$ (cross-) correlations through the halo parameters $\mathbf{p}_\times = \{\epsilon_1, \epsilon_2, M_0, \beta\}$ that are shared between the matter and pressure model.

To isolate the impact of halo parameter self-calibration, we extend the halo parameterisation with a second copy of parameters ${\mathbf{p}}'_\times$ to decouple the halo parameters for matter and pressure. This allows us to design two scenarios that fall between the fiducial 6\x2pt and 8\x2pt (6\x2pt, $y$-$y$, shear-$y$) analyses as limiting cases:
\begin{itemize}
    \item 8\x2pt \emph{without} halo parameter self-calibration: independent halo parameters for matter and pressure, i.e. $y$ field modelled by ${\mathbf{p}}'_\times$. 
    \item 8\x2pt with \emph{partial} halo parameter self-calibration: $y$ field in shear-$y$ correlations shares ${\mathbf{p}}_\times$ with the matter field, while the $y$ field in $y$-$y$ uses ${\mathbf{p}}'_\times$.
\end{itemize}

\begin{figure}
    \centering
    \includegraphics[width=3.25in]{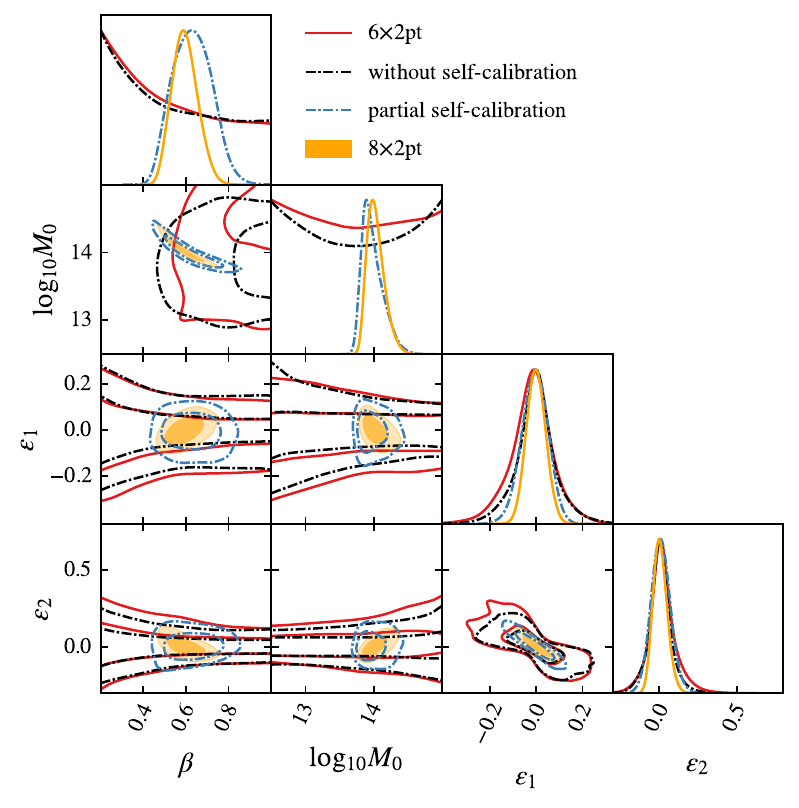}
    \caption{\label{fig:nocalib_4param}The effect of halo parameter self-calibration on halo parameters $\beta, M_0,\epsilon_1,\epsilon_2$ on 6\x2pt and 8\x2pt analyses with $\ell_{\rm max}^{\rm shear}=8000$. Starting from the fiducial 8\x2pt analysis (orange shaded) with a unified set of halo parameters for both $\delta_\mathrm{m}$ and $y$ fields, we decouple halo parameters for matter and pressure to reduce the halo parameter self-calibration partially (blue dashed-dotted) and turning it off completely (black dashed-dotted) as detailed in Sect~\ref{ssec:calib}. The 8\x2pt analysis without self-calibration approaches the constraining power of a 6\x2pt analysis (red solid). Note that the orange, blue and black contours all are 8\x2pt analyses, \ie have the same data S/N.}
\end{figure}

The corresponding constraints on ${\mathbf{p}}_\times$ are shown in Fig.~\ref{fig:nocalib_4param}. Without self-calibration between matter field and $y$ field (black dashed-dotted line), the 8\x2pt constraints are very similar to those of a 6\x2pt analysis (red line), indicating that there is no significant information gain from including tSZ. Partial self-calibration noticeably improves constraints on the halo parameters (blue dashed-dotted line), but falls short of the fiducial 8\x2pt analysis with full self-calibration (orange shaded contours). 

These results demonstrate the importance of joint parameterisation for matter and pressure to maximise the cosmology constraining power from tSZ cross-correlation analyses.

\section{Summary and Discussion}
\label{sec:discussion}
The forthcoming overlap of large galaxy photometric surveys and CMB experiments offers the opportunity to jointly analyse these datasets, thereby enhancing constraints on cosmological physics. In this paper, we present the first joint ``10\x2pt'' analysis of the galaxy position and lensing shear fields from galaxy surveys like LSST, and the reconstructed CMB lensing convergence and Compton-$y$ fields from CMB experiments like SO. These simulated analysis are based on MCMC chains with a multi-component halo model, non-Gaussian covariances, and extensive modelling of observational (shear calibration and photo-$z$ uncertainties) and astrophysical (galaxy bias, intrinsic alignment, halo) systematics.

As an example, we simulate the LSST Y6 + SO Y5 joint analysis. For this specific analysis setup, our main findings are summarised as follows:
\newcounter{summarylist}
\begin{list}{(\arabic{summarylist})\ }{\usecounter{summarylist}}
    \item Within $\Lambda$CDM, the 10\x2pt constraints of $\Omega_{\rm{m}}$ and $S_8$ improve by around 70\% in Figure-of-Merit (FoM) from 3\x2pt, or around 30\% from 6\x2pt (Sect.~\ref{ssec:constraints}).
    
    \item 8\x2pt (excluding galaxy-$y$ and CMB lensing-$y$ correlations) analysis results in cosmology constraints similar to those from 10\x2pt, suggesting that shear-$y$ and $y$-$y$ correlations are the most valuable additions when the tSZ information is included (Sect.~\ref{ssec:constraints}).
    
    \item The small-scale modelling is limited by the halo model uncertainties. By eliminating those uncertainties, one may see another $\sim$50\% improvement in FoM of the fiducial 10\x2pt. These uncertainties are mostly attributed to parameters $\epsilon_1,\epsilon_2,\Gamma$, while $M_0$ and $\beta$ contribute slightly less (Sect.~\ref{ssec:halo-impact}).

    \item We demonstrate that the Compton-$y$ field provides crucial halo self-calibration, which reduces the uncertainties in the small-scale matter power spectrum, hence indirectly improving the cosmological constraints (Sect.~\ref{ssec:calib}).
\end{list}
\begin{figure}
    \centering
    \includegraphics[width=3.25in]{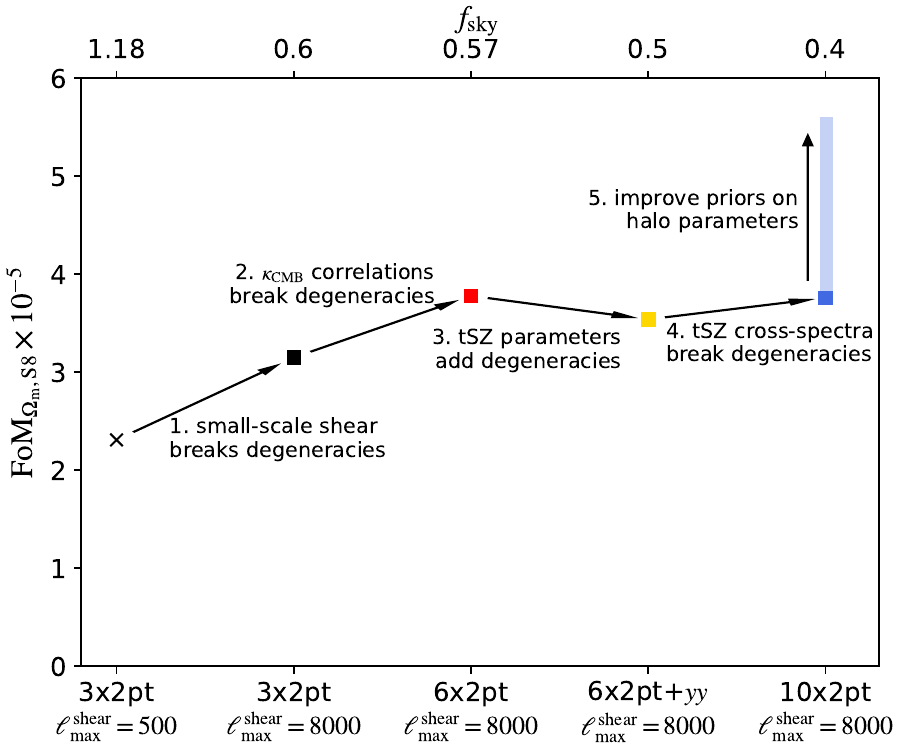}
\caption{The Figure-of-Merit (FoM) of different \emph{S/N matched-} analyses. The survey area required for each analysis to achieve the same S/N as the 10\x2pt, $\ell_{\rm{max}}^{\rm{shear}}=8000$ analysis with $f_{\rm{sky}} =0.4$ is listed on the upper horizontal axis. See text for details.}
    \label{fig:fom_fksy}
\end{figure}
More generally, the gain in constraining power from LSS\x CMB joint analyses is caused by both \emph{(i)} the increase in S/N from additional measurements and \emph{(ii)} the improved conversion of S/N to parameter constraints due to enhanced self-calibration of nuisance parameters. To illustrate the latter point, we consider different multi-probe analyses that match the S/N of our 10\x2pt, $\ell_{\rm{max}}^{\rm{shear}}=8000$ analysis, which amounts to artificially rescaling the survey area for different analyses. For example, a conservative, $\ell_{\rm{max}}^{\rm{shear}}=500$ 3\x2pt analysis would require an (impossible) survey area of $f_{\rm{sky}} =1.18$ to reach the same S/N as the 10\x2pt, $\ell_{\rm{max}}^{\rm{shear}}=8000$ analysis with $f_{\rm{sky}} =0.4$. 

Figure~\ref{fig:fom_fksy} shows the constraining power of these S/N-matched analyses, with the upper horizontal axis indicating the rescaled survey area. The different 3\x2pt analyses and 6\x2pt analysis employ the same set of model parameters; adding qualitatively different information in steps 1 and 2 breaks degeneracies between nuisance parameters thus improves the constraining power (at fixed S/N). Adding the tSZ auto-power spectrum to the 6\x2pt analysis (step 3) requires additional nuisance parameters, that are partially degenerate with other model parameters, which at fixed S/N leads to a reduction in constraining power (\ie, in this case the tSZ model parameter are constrained at the expense of the cosmological model). Including all tSZ cross-correlations in the 10\x2pt analysis then enables improved self-calibration (step 4). We illustrate the potential gain from improved priors on halo model parameters in step 5, where the extent of the blue vertical bar corresponds to the ``fix halo’’ analysis in Fig.~\ref{fig:fixhalo} as a limiting case. We discussed in the previous sections how fixing only three of the halo parameters recovers a large fraction of this gain. We note that the similar constraining power of the S/N-matched 6\x2pt and 10\x2pt analyses is likely a coincidence, and different priors on the pressure halo model parameters would change the relative constraining power of these two analyses.
The survey area required for these different S/N-matched analyses further illustrates the potential of 10\x2pt analyses and the impact of further research into joint modelling approaches: even a full-sky LSS survey is insufficient to reach comparable S/N with 3\x2pt analyses restricted to moderately non-linear scales ($\ell_{\rm{max}}^{\rm{shear}}=500$). With $\ell_{\rm{max}}^{\rm{shear}}=8000$, 3\x2pt and 6\x2pt analyses would still require $f_{\rm{sky}}\sim0.6$ surveys. The latter is in principle achievable from the ground with multiple facilities, but the associated financial and environmental costs motivate the optimisation of analysis strategies and information extraction. While the accuracy of current modelling prescriptions (like HMx) are insufficient for ambitious 10\x2pt analyses, these forecasts demonstrate the outsized impact of joint analyses on constraining power and motivate future research on joint multi-probe modelling and  multi-wavelength calibration of halo properties.

It is important to stress that beyond the increase of constraining power showcased in this paper, multi-probe analysis programs will be essential to assess model accuracy by enabling consistency checks between different data splits. 
For example, one can measure halo model parameters separately from two subsets of the 10\x2pt data, namely $y\times\{y,\delta_{\rm{g}},\kappa_{\rm{CMB}}\}$, 
which measures halo model parameters from $y$ and is most sensitive to cluster-sized halos, and $\kappa_{\rm{g}}\times\{\kappa_{\rm{g}},\delta_{\rm{g}},\kappa_{\rm{CMB}}\}$, 
which measures halo model parameters from the matter power spectrum and is most sensitive to group-sized halos. Any inconsistency in these two sets of halo model parameter constraints would indicate incompleteness of the model, e.g., in the parameterisation for the halo mass-dependence of baryonic effects.

While these conclusion inherently depend on the survey and analysis choices, we demonstrate the potential of future joint analyses of galaxy surveys and CMB experiments. In the companion paper (Fang et al, in prep), we apply the 10\x2pt methodology and analysis framework to Roman Space Telescope and SO as well as CMB-S4, with various survey and analysis choices. 


\section*{Acknowledgements}
We thank Shivam Pandey, Emmanuel Schaan, Blake Sherwin and Chun-Hao To for helpful discussions. We thank the anonymous referee for helpful comments that improved this manuscript.

XF is supported by the BCCP fellowship at the Berkeley Center for Cosmological Physics. EK is supported in part by Department of Energy grant DE-SC0020247, the David and Lucile Packard Foundation and an Alfred P. Sloan Research Fellowship. TE is supported in part by Department of Energy grant DE-SC0020215. SF is supported by the Physics Division of Lawrence Berkeley National Laboratory and by the U.S. Department of Energy (DOE), Office of Science, under contract No.
DE-AC02-05CH11231. KB, EA and YD acknowledge support by the CNES, the French national space agency. EA and PRS are supported in part by a PhD Joint Programme between the CNRS and the University of Arizona. Calculations in this paper use High Performance Computing (HPC) resources supported by the University of Arizona TRIF, UITS, and RDI and maintained by the UA Research Technologies department.

\section*{Data Availability}
The data underlying this paper will be shared on reasonable request
to the corresponding author.

\bibliographystyle{mnras}
\bibliography{references.bib}
\appendix
\section{Two-point function modelling}\label{ssec:2pt}
This appendix summarises the computation of angular (cross) power spectra for 10\x2pt analyses. We use capital Roman subscripts to denote observables, $A,B\in\lbrace\delta_{\rm g}, \kappa_{\rm g},\kappa_{\rm CMB},y\rbrace$. We assume General Relativity (GR) and a flat $\Lambda$CDM cosmology throughout.

\subsection{Angular power spectra}\label{sssec:probe-modeling}
Adopting the Limber approximation \citep{1953ApJ...117..134L} (but see \citealp{2008PhRvD..78l3506L,2020JCAP...05..010F} for potential impact in current and near future surveys), we write the angular power spectrum between redshift bin $i$ of observable $A$ and redshift bin $j$ of observable $B$ at Fourier mode $\ell$ as
\begin{equation}
C_{AB}^{ij}(\ell)=\int d\chi\frac{q_A^i(\chi)q_B^j(\chi)}{\chi^2}P_{AB}\left(\frac{\ell+1/2}{\chi},z(\chi)\right)~,
\end{equation}
where $\chi$ is the comoving distance, $P_{AB}(k,z)$ is the 3D cross-probe power spectra, $q_{A}^i(\chi),q_{B}^j(\chi)$ are weight functions of the observables $A,B$ given by
\begin{align}
    &q_{\delta_{\rm g}}^i(\chi) = \frac{n_{\rm lens}^i(z(\chi))}{\bar{n}^i_{\rm lens}}\frac{dz}{d\chi}~,\\
    &q_{\kappa_{\rm g}}^i(\chi) = \frac{3H_0^2\Omega_{\rm{m}}}{2c^2}\frac{\chi}{a(\chi)}\int_{\chi_{\rm min}^i}^{\chi_{\rm max}^i}\,d\chi'\frac{n_{\rm source}^i(z(\chi'))}{\bar{n}_{\rm source}^i}\frac{dz}{d\chi'}\frac{\chi'-\chi}{\chi'}~, \\
    &q_{\kappa_{\rm CMB}}(\chi) = \frac{3H_0^2\Omega_{\rm{m}}}{2c^2}\frac{\chi}{a(\chi)}\frac{\chi^*-\chi}{\chi^*}~, \\
    &q_y(\chi) = \frac{\sigma_{\rm T}}{m_e c^2}\frac{1}{[a(\chi)]^2}~,
\end{align}
where $\chi_{\rm min/max}^i$ is the minimum / maximum comoving distance of the redshift bin $i$, $a(\chi)$ is the scale factor, $\Omega_{\rm m}$ is the matter density fraction at present, $H_0$ is the Hubble constant, $c$ is the speed of light, $\chi^*$ is the comoving distance to the surface of last scattering, $\sigma_{\rm T}$ is the Thomson scattering cross section, and $m_e$ is the electron mass. Note that the weight functions of $y$ and $\kappa_{\rm CMB}$ do not depend on redshift bins.

We relate the 3D cross-power spectra $P_{AB}(k,z) = P_{BA}(k,z)$ to the nonlinear matter power spectrum $P_{\delta\delta}(k,z)$, matter-electron pressure power spectrum $P_{\delta P_{\rm e}}(k,z)$, and electron pressure power spectrum $P_{P_{\rm e}P_{\rm e}}(k,z)$, where $\delta$ is the nonlinear matter density contrast and $P_{\rm e}$ is the electron pressure:
\begin{align}
    &P_{\delta_{\rm g}B}(k,z) = b_{\rm g}(z)P_{\delta B}(k,z)~,\\
    &P_{\kappa_{\rm g}B}(k,z) = P_{\kappa_{\rm CMB}B}(k,z) = P_{\delta B}(k,z)~,\\
    &P_{yB}(k,z)= P_{P_{\rm e} B}(k,z)~,\label{eq:y-Pe}
\end{align}
where we have assumed that the galaxy density contrast is proportional to the nonlinear matter density contrast, weighted by an effective galaxy bias parameter $b_{\rm g}(z)$. We impose conservative scale cuts on $\delta_{\rm g}$-related probes to ensure the validity of this assumption in our analysis. Modelling beyond linear galaxy bias would be required to robustly extract cosmological information from smaller scales \citep[\eg,][]{2017MNRAS.470.2100K,2017JCAP...08..009M,2020PhRvD.102l3522P,2022PhRvD.106d3520P, 2021JCAP...12..028K,2021MNRAS.501.1481H,2021MNRAS.501.6181K}. Due to Eq.~(\ref{eq:y-Pe}), we will use $P_{yB}$ to represent $P_{P_eB}$, which simplifies the subscripts. Thus, $P_{\delta P_e}\equiv P_{\delta y}$ and $P_{P_e P_e}\equiv P_{yy}$.

We calculate the 3D power spectra $P_{AB}$ ($A,B\in\lbrace \delta, y\rbrace$) using a halo model implementation, which is based on the {\py HMx} parameterisation from \citet{2020A&A...641A.130M}.

\subsubsection{Halo Model}\label{sssec:halo-model}
The halo model \citep{2000MNRAS.318..203S,Ma_HM,Peacock_HM,2002PhR...372....1C} describes the distribution of matter, and biased tracers, based on the assumption that all matter is distributed within halos. Our implementation of the gas component largely follows the HMx code, while the other ingredients of the halo model are adapted to the existing CosmoLike implementation \citep{2017MNRAS.470.2100K}, including the halo definition, critical density contrast, and halo bias and mass function. While more accurate models and consistent implementations will be required for data analyses, we do not expect these modelling approximations to significantly impact the exploration of information content in different probe combinations presented here.

The 3D power spectrum of two fields $A$ and $B$ can then be written as the sum of a 2-halo (2h) and 1-halo (1h) term,
\begin{align}
    P_{AB}(k,z) &= P^{\rm 2h}_{AB}(k,z)+P^{\rm 1h}_{AB}(k,z)\nonumber\\
    &= P_{\rm lin}(k,z)I_{A}^1(k,z)I_{B}^1(k,z) + I_{AB}^0(k,k,z)~,
\end{align}
where $P_{\rm lin}(k,z)$ is the linear matter power, and
\begin{align}
    I_{A}^\alpha(k,z) &= \langle b_{h,\alpha}(M,z) \tilde{u}_{A}(k,M,z)\rangle_{\scriptscriptstyle M}~,\\
    I_{AB}^\alpha(k,k',z) &= \langle b_{h,\alpha}(M,z) \tilde{u}_{A}(k,M,z)\tilde{u}_{B}(k',M,z)\rangle_{\scriptscriptstyle M}~,
\end{align}
where $b_{h,\alpha}$ is the $\alpha$-th order halo biasing, with $b_{h,0}=1$, and where we neglect higher order biasing, \ie, $b_{h,\geq 2}=0$. The functions $\tilde{u}_X(k,M,z)$ ($X\in\lbrace \delta,y\rbrace$) are related to the Fourier transforms of the radial profiles of matter density $\rho_{\rm m}(M,r,z)$ and electron pressure $P_{\rm e}(M,r,z)$ within a halo of mass $M$, and where we have used a shorthand notation for the halo mass function-weighted average of quantity $X$,
\begin{equation}
    \langle X(M)\rangle_{\scriptscriptstyle M} \equiv \int_{M_{\rm min}}^{M_{\rm max}} dM\frac{dn(M)}{dM}X(M)~,
\end{equation}
where $M$ is the halo mass and $dn(M)/dM$ is the halo mass function.

This standard halo model calculation leads to unphysical behaviour of the 1h term as $k\rightarrow 0$. We follow the treatment in the public halo model code {\py HMcode-2020} \citep[Eq.~17 of][]{2021MNRAS.502.1401M} for all 1h terms at low-$k$,
\begin{equation}
    P_{AB}^{\rm 1h}(k,z)\rightarrow P_{AB}^{\rm 1h}(k,z)\frac{(k/k_*)^4}{1+(k/k_*)^4}~,
\end{equation}
and take $k_*$ as the fitted functional form in their Tab.~2.

\subsection{Systematics}\label{ssec:sys}
Systematic uncertainties are parameterised through nuisance parameters, whose fiducial values and priors are summarised in Tab.~\ref{tab:params}. The nuisance parameters included in our simulated analysis are very similar to those used in the joint Rubin Observatory - Simons Observatory forecast in \cite{2022MNRAS.509.5721F} and the Rubin Observatory - Roman Space Telescope forecast in \cite{2021MNRAS.507.1514E}, except that here we do not use a principal component analysis approach \citep{2019MNRAS.488.1652H,2021MNRAS.502.6010H} for the baryonic feedback. Instead, we have treated the baryonic feedback in a consistent halo model as described in the previous subsection. The systematic effects are summarised as follows:

\paragraph*{Photometric redshift uncertainties}
We parameterise photometric redshift uncertainties by a Gaussian scatter $\sigma_{z,x}$ for lens and source sample each, and a shift parameter $\Delta_{z,x}^i$ for each redshift bin $i$ of the lens and source samples. The binned true redshift distribution $n_x^i(z)$ is related to the binned photometric redshift distribution $n_x^i(z_{\rm ph})$ (Eq.~\ref{eq:nz-true}) by
\begin{equation}
    n_x^i(z) = \int_{z_{{\rm min},x}^i}^{z_{{\rm max},x}^i}\frac{dz_{\rm ph}\,n_x^i(z_{\rm ph})}{\sqrt{2\pi}\sigma_{z,x}(1+z_{\rm ph})}\exp\left[-\frac{(z-z_{\rm ph}-\Delta_{z,x}^i)^2}{2[\sigma_{z,x}(1+z_{\rm ph})]^2}\right]~.
\label{eq:binned-nz-true}
\end{equation}
We set the fiducial values of $\Delta_{z,x}^i$ as zero, $\sigma_{z,{\rm lens}}$ as 0.03, and $\sigma_{z,{\rm source}}$ as 0.05, following the DESC-SRD. The resulting distributions are shown in Fig.~\ref{fig:nz}. In total, we have 22 photo-$z$ parameters (10 shift parameters and 1 scatter parameter for lens and source sample each). In this paper, these parameters are analytically marginalised over using Gaussian priors, as described in Sect.~\ref{subsec:analytic-margin}.

For the source samples, the $\Delta_{z,{\rm source}}^i$ and $\sigma_{z,{\rm source}}$ priors for LSST Y6 are chosen to be the same as the Y10 requirements given in DESC-SRD. For the lens samples, we choose these priors to be same as the corresponding source sample priors\footnote{This choice follows \cite{2022MNRAS.509.5721F} and is justified in their footnote 4.}. 

\paragraph*{Linear galaxy bias}
We assume one linear bias parameter $b_{\rm g}^i$ per lens redshift bin, whose fiducial values follow the simple relation as given in the DESC-SRD: $b_{\rm g}^i=0.95/G(\bar{z}^i)$, where $G(z)$ is the growth function. We independently marginalise over the 10 linear bias parameters with a conservative flat prior $[0.4,3.0]$. We note that this model may be oversimplified for going into $k_{\rm max}=$0.3$h/$Mpc. More complex models may lead to less constraining power of the galaxy survey and enhance the importance of the secondary CMB information.

\paragraph*{Multiplicative shear calibration}
We assume one parameter $m^i$ per source redshift bin, which affects $\kappa_{\rm g}$-$X$ ($X\in\lbrace \kappa_{\rm g}, \delta_{\rm g},\kappa_{\rm CMB},y \rbrace$) via
\begin{equation}
    q_{\kappa_{\rm g}}^i\rightarrow (1+m^i)q_{\kappa_{\rm g}}^i
\end{equation}
The total of 10 $m^i$ parameters are independently marginalised over with Gaussian priors shown in Tab.~\ref{tab:params}. The priors are chosen to be the same as the LSST Y10 requirements given in DESC-SRD. In this paper, we analytically marginalise these parameters as described in Sect.~\ref{subsec:analytic-margin}.

\paragraph*{Intrinsic alignment (IA)}
We adopt the ``nonlinear linear alignment'' model \citep{2004PhRvD..70f3526H,2007NJPh....9..444B}, which considers only the ``linear'' response of the elliptical (red) galaxies' shapes to the tidal field sourced by the underlying ``nonlinear'' matter density field.
Our implementation follows \cite{2016MNRAS.456..207K} for cosmic shear, \cite{2017MNRAS.470.2100K} for galaxy-galaxy lensing, \cite{2022MNRAS.509.5721F} for $\kappa_{\rm g}$-$\kappa_{\rm CMB}$ power spectra, and extend it to $\kappa_{\rm g}$-$y$ power spectra. Using the notation in Sect.~\ref{sssec:probe-modeling}, we can encapsulate the effect as
\begin{equation}
    q_{\kappa_{\rm g}}^i\rightarrow q_{\kappa_{\rm g}}^i + q_I^i~,
\end{equation}
where
\begin{equation}
    q_I^i = -A(z)\frac{n_{\rm source}^i(z(\chi))}{\bar{n}_{\rm source}^i}\frac{dz}{d\chi}~.
\end{equation}
$A(z)$ is the IA amplitude at a given redshift $z$, computed by
\begin{equation}
    A(z)=\frac{C_1\rho_{\rm cr}}{G(z)}A_{\rm IA} \left(\frac{1+z}{1+z_0}\right)^{\eta_{\rm IA}}~,
\end{equation}
with pivot redshift $z_0=0.3$, and with $C_1\rho_{\rm cr}=0.0134$ derived from SuperCOSMOS observations \citep{2004PhRvD..70f3526H,2007NJPh....9..444B}. The fiducial values and priors for the nuisance parameters $A_{\rm IA}$ and $\eta_{\rm IA}$ are given in Tab.~\ref{tab:params}. We neglect luminosity dependence and additional uncertainties in the luminosity function, which can be significant and are discussed in \cite{2016MNRAS.456..207K}.

Together with the multiplicative shear calibration, $q_{\kappa_{\rm g}}^i$ is altered as
\begin{equation}
    q_{\kappa_{\rm g}}^i(\chi)\rightarrow (1+m^i)[q_{\kappa_{\rm g}}^i(\chi)+q_I^i(\chi)]~.
\end{equation}

We do not consider higher-order tidal alignment, tidal torquing models \citep[see \eg,][]{2015JCAP...08..015B,2019PhRvD.100j3506B}, or effective field theory models \citep{2020JCAP...01..025V}, or more complicated IA modelling as a function of galaxy colour \citep{2019MNRAS.489.5453S}, or IA halo model \citep{2021MNRAS.501.2983F}. Similar to non-linear galaxy bias models, the degrees of freedom that are opened up by these models may degrade the constraining power of the galaxy survey and enhance the importance of the information carried by secondary CMB effects.

\section{Analytic Covariances}\label{ssec:cov}
The Fourier 10\x2pt covariance matrix $\bmat{C}(C(\ell_1),C(\ell_2))$ includes the Gaussian part $\bmat{C}^{\rm G}$ \citep{2004PhRvD..70d3009H}, the non-Gaussian part from connected 4-point functions (in short, ``connected non-Gaussian'' or cNG) in the absence of survey window effect $\bmat{C}^{\rm cNG}$ \citep[\eg,][]{2002PhR...372....1C,2009MNRAS.395.2065T}, and the super-sample covariance (SSC) $\bmat{C}^{\rm SSC}$ \citep{2013PhRvD..87l3504T}, \ie, $\bmat{C}= \bmat{C}^{\rm G}+ \bmat{C}^{\rm cNG} + \bmat{C}^{\rm SSC}~$.

\paragraph*{Gaussian covariance}
We extending the modelling and implementation described in \cite{2017MNRAS.470.2100K,2017PhRvD..95l3512S,2022MNRAS.509.5721F} to include the $y$-related probes. The Gaussian covariance includes the probe-specific shot noise terms, \ie,
\begin{align}
    &\bmat{C}^{\rm G}(C_{\rm AB}^{ij}(\ell_1),C_{\rm CD}^{kl}(\ell_2)) \nonumber\\
    &=\frac{\delta_{\ell_1\ell_2}}{f_{\rm sky}(2\ell_1+1)\Delta\ell_1}\left[\hat{C}_{\rm AC}^{ik}(\ell_1)\hat{C}_{\rm BD}^{jl}(\ell_2)+\hat{C}_{\rm AD}^{il}(\ell_1)\hat{C}_{\rm BC}^{jk}(\ell_2)\right]~,
\end{align}
where $\delta_{ij}$ is the Kronecker delta function, $f_{\rm sky}=\Omega_{\rm s}/(4\pi)$ is the survey's sky coverage fraction, $\Delta\ell$ is the $\ell$ bin width, $\hat{C}_{\rm AC}^{ik}(\ell_1)=C_{\rm AC}^{ik}(\ell_1)+\delta_{ik}\delta_{\rm AC}N_A^i(\ell_1)$. The noise term $N_A^i(\ell)$ are given by
\begin{equation}
    N_{\kappa_{\rm g}}^i(\ell)=\sigma_{\epsilon}^2/\bar{n}_{\rm source}^i~,~~N_{\delta_{\rm g}}^i(\ell)=1/\bar{n}_{\rm lens}^i~,
\end{equation}
where $\sigma_\epsilon=0.26$ is the shape noise per component given by the DESC-SRD.
$N_{\kappa_{\rm CMB}}(\ell)$, $N_{y}(\ell)$ are the CMB lensing reconstruction noise and the Compton-$y$ reconstruction noise.

The Simons Observatory (SO) noise models are based on component separated maps for the ``goal'' sensitivity (\texttt{SENS-2}), and include the expected level of residual foregrounds as well as atmospheric noise. The noise for the Compton-$y$ (tSZ) map\footnote{\url{https://github.com/simonsobs/so_noise_models/blob/master/LAT_comp_sep_noise/v3.1.0/SO_LAT_Nell_T_atmv1_goal_fsky0p4_ILC_tSZ.txt}}, is obtained from the standard ``Internal Linear Combination'' (ILC, \texttt{deproj0}) without any additional deprojection on $40\%$ of the sky. For the noise properties of the CMB lensing map\footnote{\url{https://github.com/simonsobs/so_noise_models/blob/master/LAT_lensing_noise/lensing_v3_1_1/nlkk_v3_1_0_deproj0_SENS2_fsky0p4_it_lT30-3000_lP30-5000.dat}} we take the same sensitivity and area as the tSZ map, with the minimum variance ILC, including both temperature and polarisation data (with $\ell_{\rm min,T}=\ell_{\rm min,P}=30$, $\ell_{\rm max,T}=3000$, and $\ell_{\rm max,P}=5000$) and no additional deprojection (\texttt{deproj0}). The lensing noise includes a small improvement expected from iterative lensing reconstruction. These noise curves were obtained with the methodology explained in Section 2 of \cite{2019JCAP...02..056A} and are publicly available on GitHub at the address linked in the footnotes below.

\paragraph*{Connected non-Gaussian (cNG) covariance}
The cNG term is computed from the projection of the trispectra. We neglect the mode-coupling due to the survey window and adopt the Limber approximation,
\begin{align}
    &\bmat{C}^{\rm cNG}(C_{\rm AB}^{ij}(\ell_1),C_{\rm CD}^{kl}(\ell_2)) = \frac{1}{4\pi f_{\rm sky}}\int_{\bm{\ell}\in\ell_1}\frac{d^2\bm{\ell}}{A(\ell_1)}\int_{\bm{\ell}'\in\ell_2}\frac{d^2\bm{\ell}'}{A(\ell_2)}\nonumber\\
    &\times \int d\chi\frac{q_{\rm A}^i(\chi) q_{\rm B}^j(\chi)q_{\rm C}^k(\chi)q_{\rm D}^l(\chi)}{\chi^6} T_{\scriptscriptstyle\rm ABCD}^{ijkl}\left(\frac{\bm{\ell}}{\chi},-\frac{\bm{\ell}}{\chi},\frac{\bm{\ell}'}{\chi},-\frac{\bm{\ell}'}{\chi};z(\chi)\right).
\end{align}
In the case where none of A,B,C,D fields is $y$ field, we approximate the trispectrum as
\begin{align}
    T_{\scriptscriptstyle\rm ABCD}^{ijkl}&(\bv{k},-\bv{k},\bv{k}',-\bv{k}';z)\simeq T_{\scriptscriptstyle\rm ABCD}^{\rm 1h}(k,k,k',k';z)\nonumber\\
    &+b^i_{\rm A}(z)b^j_{\rm B}(z)b^k_{\rm C}(z)b^l_{\rm D}(z)T_{\rm m}^{\rm 2h+3h+4h}(\bv{k},-\bv{k},\bv{k}',-\bv{k}';z)~,
\end{align}
where the bias $b_A^i(z)$ is equal to the galaxy bias $b_{\rm g}^i(z)$ if ${\rm A}=\delta_{\rm g}$, and $b_A^i(z)=1$ otherwise. For (cross-) covariances involving at least one $y$ field, we only consider the 1-halo contribution \citep{2002MNRAS.336.1256K,2018MNRAS.480.3928M}
\begin{align}
    T_{\scriptscriptstyle\rm ABCy}(\bv{k},-\bv{k},\bv{k}',-\bv{k}';z) = \langle &\tilde{u}_{\scriptscriptstyle\rm A}(k,M,z)\tilde{u}_{\scriptscriptstyle\rm B}(k,M,z)\nonumber\\
    &\tilde{u}_{\scriptscriptstyle\rm C}(k',M,z)\tilde{u}_{\scriptscriptstyle\rm y}(k',M,z)\rangle_{\scriptscriptstyle M}.
\end{align}

\paragraph*{Super-sample covariance (SSC)}
We extend the {\py CosmoCov} SSC implementation to include $y$ cross-correlations, following the formalism of \citep{2013PhRvD..87l3504T} applied to multi-probe observables \citep{2017MNRAS.470.2100K} and assuming a polar cap survey footprint to evaluate the variance of super-survey density modes.

\begin{figure}
    \centering
    \includegraphics[width=3.25in]{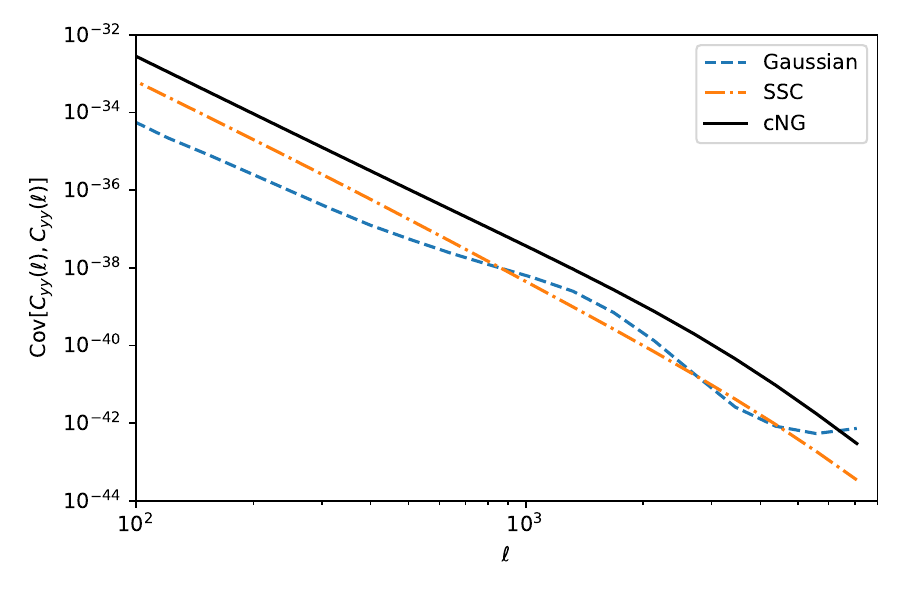}
    \caption{The diagonal elements of the tSZ power spectrum covariance. The blue, orange and green lines show the Gaussian covariance, the super-sample covariance (SSC), the connected non-Gaussian (cNG) covariance, respectively, with cNG dominates over other two components for the entire range of angular scales considered. The cNG component computed here only includes the 1-halo contributions.}
    \label{fig:cov_yy_diag}
\end{figure}

\begin{figure*}
    \centering
    \includegraphics[width=6.75in]{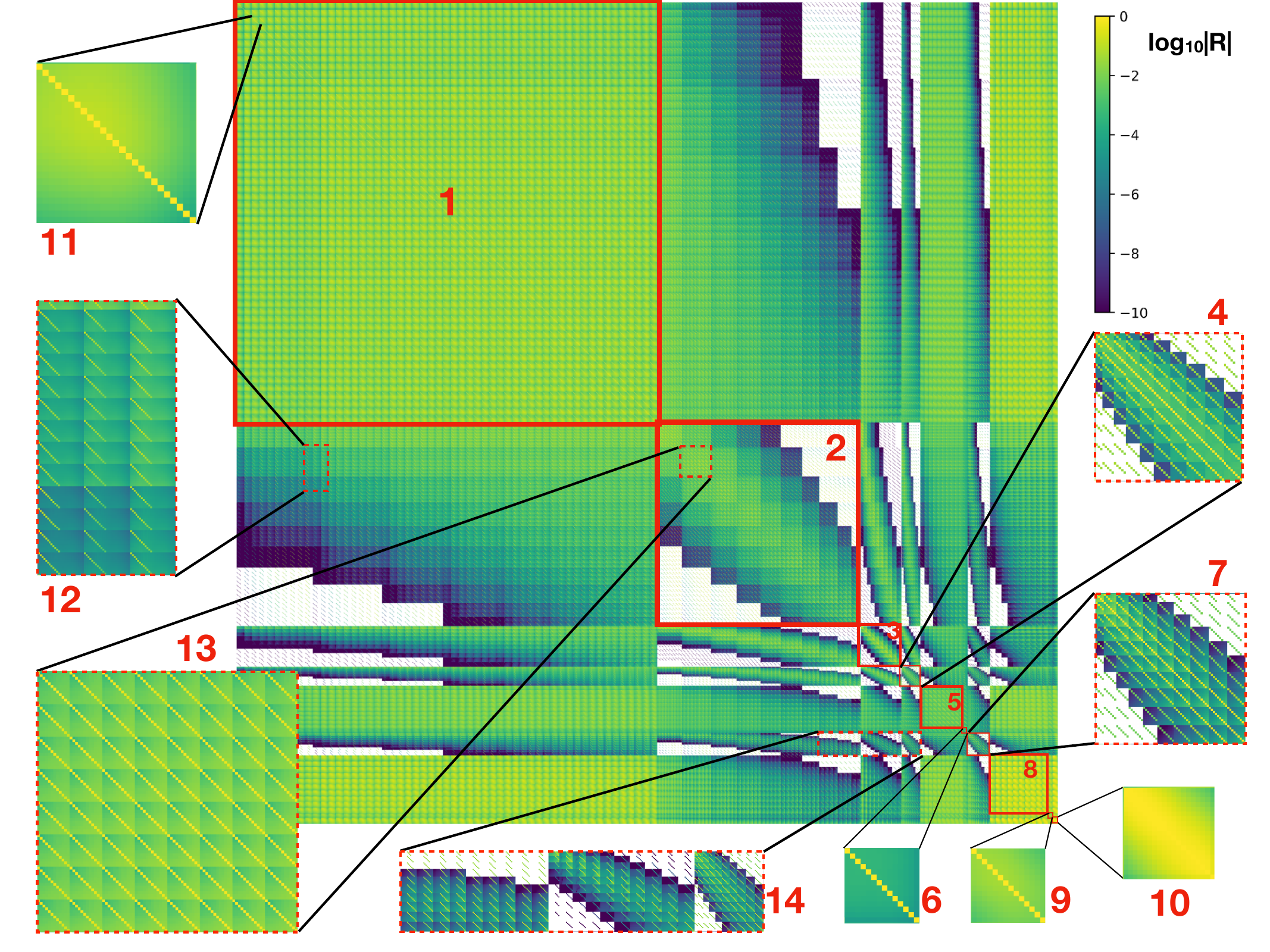}
    \caption{The 10\x2pt correlation matrix for LSST Y6 \x SO Y5, plotted as $\log_{10}|\bmat{R}|$, where $\bmat{R}=[\rm{diag}(\bmat{C})]^{-1/2}\bmat{C}[\rm{diag}(\bmat{C})]^{-1/2}$. We have highlighted some parts of the matrix to illustrate the correlation structure: (1) depicts the cosmic shear ($\kappa_{\rm g}$-$\kappa_{\rm g}$) correlation matrix, comprised of 55 tomographic combinations of source bins, each with 25 Fourier $\ell$-bins. (11) shows one of the tomographic combinations, and the individual $\ell_1, \ell_2$ elements are clearly visible. (2) is the galaxy-galaxy lensing ($\kappa_{\rm g}$-$\delta_{\rm g}$) tomography correlation with (13) being the galaxy-galaxy combinations of the 2nd lens bin with all the non-overlapping source bins at higher redshifts. (12) zooms into the matrix of a part of the correlation between cosmic shear and galaxy-galaxy lensing. (3) is the clustering ($\delta_{\rm g}$-$\delta_{\rm g}$) auto-probe matrix with 10 tomographic bins. (4) corresponds to the galaxy-CMB convergence ($\delta_{\rm g}$-$\kappa_{\rm CMB}$) combinations auto-probe matrix, which is comprised of 10 lens sample redshift bins. (5) is the auto-probe correlation of the $\kappa_{\rm g}$-$\kappa_{\rm CMB}$ part of the data vector, which uses the 10 source sample redshift bins. (6) is the auto-probe correlation of the CMB lensing power spectrum, $\kappa_{\rm CMB}$-$\kappa_{\rm CMB}$. (7) is the auto-probe correlation of the galaxy-tSZ ($\delta_{\rm g}$-$y$) combinations, which uses the 10 lens redshift bins. (14) zooms into the correlation matrix of all 10 galaxy-tSZ combinations with the last 3 of galaxy-galaxy lensing power spectra ($\kappa_{\rm g}$-$\delta_{\rm g}$), all the 10 galaxy clustering ($\delta_{\rm g}$-$\delta_{\rm g}$), and all 10 $\delta_{\rm g}$-$y$ spectra, respectively from left to right. (8) is the auto-probe correlation of the galaxy lensing convergence-tSZ ($\kappa_{\rm g}$-$y$) combinations, which uses the 10 source redshift bins. (9) is the auto-probe correlation of the CMB lensing convergence-tSZ ($\kappa_{\rm CMB}$-$y$). (10) is the auto-probe correlation of the tSZ power spectrum ($y$-$y$).}
    \label{fig:cov}
\end{figure*}

We show the full correlation matrix $\bmat{R}$, given by $\bmat{R}=[\rm{diag}(\bmat{C})]^{-1/2}\bmat{C}[\rm{diag}(\bmat{C})]^{-1/2}$, of the 10\x2pt data vector in Fig.~\ref{fig:cov}, evaluated at the fiducial parameter values. The $y$-$y$ and shear-$y$ probes are highly correlated across tomographic bins and $\ell$-bins due to the significant contributions from the cNG term. In Fig.~\ref{fig:cov_yy_diag}, we show the Gaussian, cNG, and SSC components of the diagonal elements of the $\bmat{C}(C_{yy},C_{yy})$ covariance block. The dominance of the cNG component over the SSC and the Gaussian part is consistent with \cite{2021PhRvD.103f3501O}. As shown in their paper, applying a cluster mask for massive clusters can significantly suppress the noise level. We leave the studies of its improvement of our 10\x2pt analysis for future work.

\label{lastpage}
\end{document}